\documentclass[sigconf]{acmart}
\usepackage{etex}

\usepackage{booktabs}   
\usepackage{subcaption} 

\author{Yuan Si}
\affiliation{%
  \institution{University of Waterloo}
  \city{Waterloo}
  \country{Canada}
  }
\email{yuan.si@uwaterloo.ca}

\author{Kyle Qi}
\affiliation{%
  \institution{University of Waterloo}
  \city{Waterloo}
  \country{Canada}
  }
\email{kdqi@uwaterloo.ca}

\author{Daming Li}
\affiliation{%
  \institution{Independent Researcher}
  \country{USA}
  }
\email{damingliyale22@gmail.com}

\author{Hanyuan Shi}
\affiliation{%
  \institution{Independent Researcher}
  \country{China}
  }
\email{shihanyuan1995@gmail.com}

\author{Jialu Zhang}
\authornote{Corresponding Author}
\affiliation{%
  \institution{University of Waterloo}
  \city{Waterloo}
  \country{Canada}
  }
\email{jialu.zhang@uwaterloo.ca}

\usepackage{graphicx}
\usepackage{amsmath}

\usepackage{xspace}
\usepackage{amsfonts}
\usepackage{natbib}
\usepackage{hhline}
\usepackage{tabularx}
\usepackage{scratch3}
\usepackage{diagbox}

\usepackage{float}
\usepackage{multirow}
\usepackage{setspace}
\usepackage{epsfig}
\usepackage{url}

\usepackage{breakurl}
\usepackage{pdfpages}
\usepackage{booktabs}
\usepackage{tabularx}
\newcolumntype{Y}{>{\raggedright\arraybackslash}X}
\usepackage{tikz}
\usetikzlibrary{arrows.meta,positioning,shapes.geometric,shapes.misc,fit,backgrounds}

\usepackage{xcolor}
\usepackage{lipsum}
\usepackage{courier}
\usepackage{listings}
\usepackage{caption}
\usepackage{colortbl}
\usepackage{arydshln} 
\usepackage{makecell} 
\usepackage{tikz}
\usepackage{color}
\usepackage{rotating}
\usepackage{varwidth}
\usetikzlibrary{fit,arrows,calc,positioning,quotes}
\usepackage{pgfplots}
\usepackage{mathtools}
\usepackage{fancyvrb}
\usepackage{textcomp}
\usepackage{longtable}
\usepackage{pdflscape} 
\usepackage{threeparttable} 
\usepackage{array}     
\usepackage{stmaryrd}
\usepackage{algorithm}
\usepackage[noend]{algpseudocode}

\usepackage{alltt}

\tikzset{
  -|-/.style={
    to path={
      (\tikztostart) -| ($(\tikztostart)!#1!(\tikztotarget)$) |- (\tikztotarget)
      \tikztonodes
    }
  },
  -|-/.default=0.5,
  |-|/.style={
    to path={
      (\tikztostart) |- ($(\tikztostart)!#1!(\tikztotarget)$) -| (\tikztotarget)
      \tikztonodes
    }
  },
  |-|/.default=0.5
}

\usetikzlibrary{shapes.geometric}

\long\def\com#1{}

\newcommand{\app}{\texttt{Stitch}\xspace}
\definecolor{ForestGreen}{RGB}{34,139,34}

\newcommand{\para}[1]{\smallskip\noindent {\bf #1}}

\newcommand{\squishlist}{
   \begin{list}{$\bullet$}
    { \setlength{\itemsep}{0pt}      \setlength{\parsep}{3pt}
      \setlength{\topsep}{3pt}       \setlength{\partopsep}{0pt}
      \setlength{\leftmargin}{3.5mm} \setlength{\labelwidth}{1em}
      \setlength{\labelsep}{0.5em} }
}

\newcommand{\squishend}{
    \end{list}  }

\def\BibTeX{{\rm B\kern-.05em{\sc i\kern-.025em b}\kern-.08em
    T\kern-.1667em\lower.7ex\hbox{E}\kern-.125emX}}

\begin{document}

\title{Stitch: Step-by-step LLM Guided Tutoring for Scratch}

\begin{abstract}
	
Block-based environments such as Scratch are increasingly popular in programming education. While block syntax reduces surface errors, semantic bugs remain common and challenging for novices to resolve. Existing debugging workflows typically show the correct program directly to learners, a strategy that may fix errors but undermines the development of problem-solving skills.

We present \app, an interactive tutoring system that replaces ``showing the answer'' with step-by-step scaffolding. The system's Diff-Analyze module contrasts a student's project with a reference implementation, identifies the most critical differences, and uses a large language model to explain why these changes matter. Learners inspect highlighted blocks through a custom rendering engine, understand the explanations, and selectively apply partial fixes. This iterative process continues until the intended functionality is achieved.

We evaluate \app in an empirical study, comparing it against a state-of-the-art automated feedback generation tool for Scratch. Our key insight is that simply presenting the correct program is pedagogically ineffective. In contrast, our interactive, step-by-step guided system promotes a more effective learning experience. More broadly, what constitutes effective feedback in block-based programming remains an open question. Our evaluation provides new evidence that step-by-step tutoring significantly enhances learning outcomes, outperforming both direct-answer approaches and current automated feedback generation tools.
\end{abstract}

\maketitle

\section{Introduction}
\label{sec:intro}

Block-based programming environments such as Scratch~\cite{resnick2009scratch,maloney2010scratch} have become foundational in Computer Science education. Designed for beginners, Scratch reduces barriers to entry by enabling users to create games, animations, and interactive stories through a drag-and-drop interface. The platform has grown to more than 140 million registered users worldwide\footnote{\url{https://annualreport.scratchfoundation.org}}, including elementary school children and online learners. By eliminating textual syntax, Scratch helps novices focus on computational logic and creativity, making programming accessible to populations that might otherwise be excluded.  

However, despite these advantages, learners frequently encounter semantic errors. A sprite may not move as expected, animations may flicker, collisions may fail to register, or entire functions may remain unexecuted. Because Scratch provides no explicit error messages, these issues often manifest simply as ``the program doesn’t do what I thought it would.''~\footnote{\url{https://scratch.mit.edu/discuss/topic/460412/}}. For beginners, bug identification from such unexpected program behaviors is itself a challenge~\cite{fradrich2020commonbugs}. Online forums are filled with posts complaining ``the cat doesn’t move when I press the key'' or ``my game never ends even after all lives are gone,'' implying both the prevalence of these problems and the difficulty learners face when diagnosing them\footnote{\url{https://scratch.mit.edu/discuss/}}. There is increasing demand for scalable, automated feedback generation that helps students recognize, understand, and resolve errors.  

Recent research has attempted to fill this gap with automated debugging and feedback systems for Scratch~\cite{price2017isnap,fein2022catnip,deiner2024nuzzlebug,STRIJBOL2024101617,schweikl2025repurr}. These tools employ techniques such as step-by-step execution, breakpoints, pattern matching, or instructor-authored test suites. While valuable, they tend to rely heavily on user's manual inputs or pre-designed, handcrafted rules, limiting their ability to handle open-ended learner projects. More importantly, most existing workflows adopt a ``repair-for-the-student'' paradigm: the system detects an error, then directly shows or applies the correct fix. While this approach rapidly produces a working program, it fails from a pedagogical point of view. Without engaging the learner in the problem-solving process, it fails to provide opportunities for deeper thinking and improving debugging skills that are crucial for programming education.  

Educational theory suggests a more effective alternative: scaffolding through step-by-step guidance~\cite{vandePol2010Scaffolding}. In real classrooms, teachers do not directly provide students with the corrected code. Instead, they highlight critical differences between a learner's solution and the intended functionality, explain why these differences matter, and encourage the learner to attempt a correction themselves~\cite{vandePol2010Scaffolding,Elliott2023DebuggingPedagogy,Elliott2024DebuggingPedagogies}. This process not only leads to a working program, but enhances conceptual understanding and problem-solving skills. Inspired by this, we translate this pedagogical style into an automated system and address the following research challenge:  

\textit{What constitutes effective feedback in block-based programming, and how can AI systems provide it in ways that preserve—rather than replace—the learner’s role in problem-solving?}

We propose \textbf{Stitch}, an interactive, \textbf{St}ep-by-step LLM gu\textbf{i}ded debugging system for Scra\textbf{tch}.
Given a teacher-provided target project as a reference, \app compares the student’s project against the target and identifies the most critical differences. Rather than automatically repairing the project, the system highlights these differences and generates explanatory hints using a large language model (LLM, Gemini 2.5 Flash Lite~\footnote{\url{https://docs.cloud.google.com/vertex-ai/generative-ai/docs/models/gemini/2-5-flash-lite}}). The explanatory hints emphasize \emph{why} changes are needed, helping learners understand not just what to change but the reasoning steps behind it. \app generates feedback using visual Scratch blocks, enabling learners to view and reason about their problematic constructs in the same format they use to create their programs. Learners can then decide whether and how to apply the suggested modifications. After each revision, the system re-analyzes the project, generates new hints, and continues until the intended functionality is achieved.  

In summary, we make the following contributions in this paper:
\begin{itemize}
    \item \textbf{System design:} We design and implement \app, an LLM-based step-by-step tutoring system for Scratch that integrates code comparison, visual rendering, LLM-generated explanations, and iterative guidance generation into a unified workflow.  
    \item \textbf{Pedagogical value:} We introduce an automated guided programming tutoring mechanism with step-by-step scaffolding that simulates proven educational practices in real classrooms.  
    \item \textbf{Empirical study:} We empirically evaluate \app on real-world Scratch projects, against a state-of-the-art automated feedback generation baseline, and show that \app achieves significantly better learning experiences and outcomes.  
\end{itemize}  

Overall, our results show that our LLM-based step-by-step interactive tutoring system substantially improves learning outcomes. Students guided by \app demonstrate more engagement and improved problem-solving skills, reaching the effectiveness of experienced teachers while significantly outperforming direct-answer strategies. While this study focuses on Scratch, the findings suggest broader implications for feedback design in block-based, and potentially introductory textual programming environments, pointing towards a generalizable paradigm of AI-based tutoring in programming education.

\section{Understanding Feedback in Scratch}
\label{sec:background}

Block-based programming offers a gentle entry point into computing, but the simplicity of its syntax hides a persistent challenge: helping learners reason about program behavior. In Scratch, most bugs are semantic rather than syntactic. Sprites may move at the wrong time, variables may fail to reset, or event sequences may fire in the wrong order. Because Scratch lacks textual error messages, learners must infer the source of failure from visual behavior alone. Effective feedback must therefore bridge a uniquely wide gap—between what the program \textbf{does} and what the learner \textbf{intended} it to do.

\begin{figure*}[t!]
  \centering
  \includegraphics[width=\linewidth]{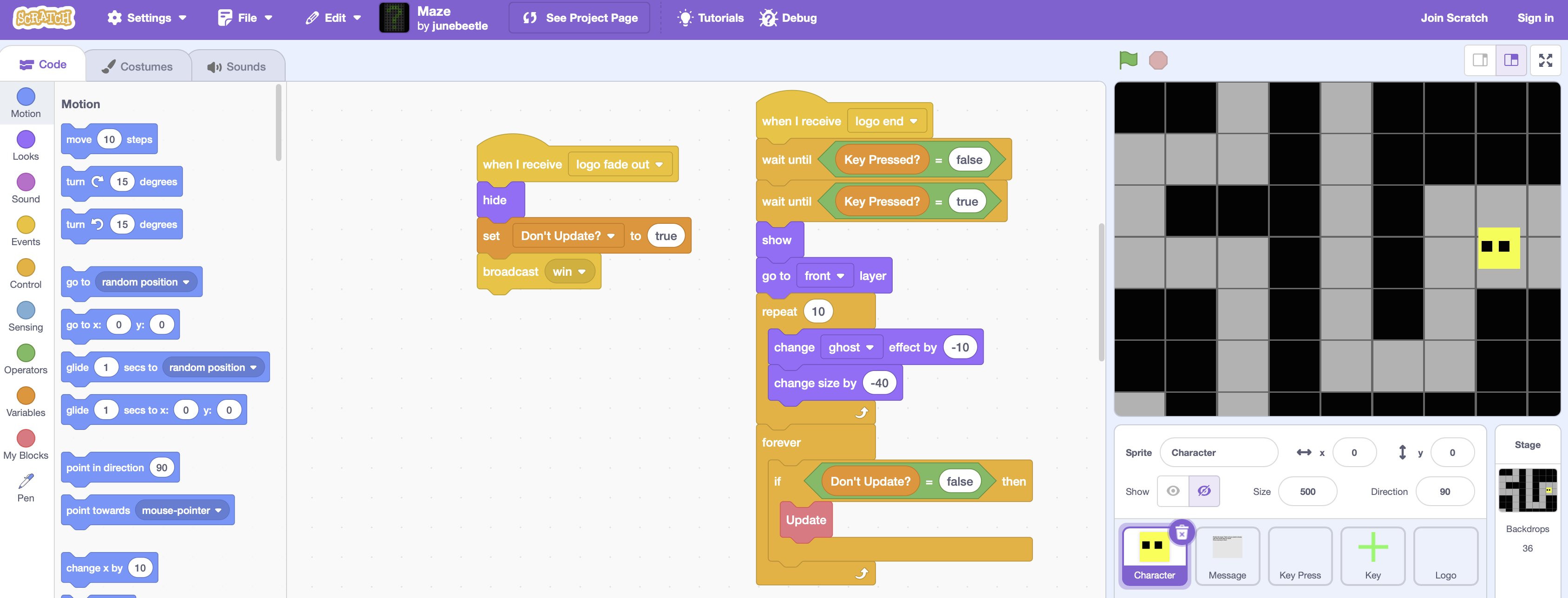}
  \caption{An example Scratch project of a maze game~\cite{resnick2009scratch}. Source: \url{https://scratch.mit.edu/projects/77278452}}
  \label{fig:GUI}
\end{figure*}

\subsection{Scratch as a Programming Environment}
Scratch~\cite{resnick2009scratch,maloney2010scratch} adopts a constructionist approach to programming, using visual blocks instead of text. Learners construct behavior by arranging color-coded blocks that represent actions such as moving sprites or responding to input. These blocks are designed with complementary shapes and strict connection rules, preventing syntax errors and allowing learners to focus on logical structure rather than textual precision.
Each project is saved as a structured JSON file that captures scripts, sprites, and media elements. This clear separation between the visual interface and the system's execution provides a solid foundation for program analysis and automation.

\para{What the User Sees.}
Scratch's Virtual Machine (VM) interprets the project's JSON model and executes scripts as event-driven threads in response to stimuli such as input, messages, or collisions. Faults are isolated at the thread level, enabling other scripts to continue running despite errors, which fosters creativity over correctness and supports learner experimentation without disruptive failures. Figure~\ref{fig:GUI} illustrates a Scratch project: the left panel offers categorized blocks for different actions, while the central canvas serves as the scripting workspace. The stage on the right visualizes program output, showing animations and sprite behaviors in real time. When something goes wrong, such as an animation stuttering, the error is visually represented on the stage rather than as a textual message. This tight coupling between action and feedback makes Scratch intuitive for novices. However, simply showing the repaired program fails to close the learning gap. Without guiding learners through the reasoning behind the changes, they may fail to grasp the underlying principles and miss out on valuable insights into debugging and program behavior. Automated feedback systems must, therefore, go beyond simply presenting the solution and instead engage learners in the process, fostering a deeper understanding of how to identify, analyze, and resolve issues.

\subsection{How Feedback is Typically Provided}
In online help forums, classrooms, and automated tools, we observe four recurring patterns of feedback. Each embodies a different philosophy of assistance, yet all share a focus on *producing correct code* rather than *cultivating understanding*.

\para{Verbal hints.} Human helpers often respond with brief, high-level suggestions such as “Make a variable for lives and decrease it when hit.”~\footnote{\url{https://scratch.mit.edu/discuss/topic/46131/}} These hints assume implicit understanding of control flow and variable scope. For novices, however, such abstraction can obscure rather than clarify. Learners recognize *what* to add, but not *how* to connect it to their existing logic~\cite{aleven_2016_help_seeking, price_aied_2017_hint_quality, wiggins_2021_hint_requests}.

\para{Direct fixes.} A frequent response is to post the complete repaired project.~\footnote{\url{https://scratch.mit.edu/discuss/topic/382741/}} This yields instant gratification—a working program—but erases the reasoning process. The student sees the *answer* but not the *path* that leads there~\cite{chi_1994_self_explanations}.

\para{Static comparisons.} Some instructors contrast a student’s project with a reference solution via code diffs or behavioral checklists.~\footnote{\url{https://scratch.mit.edu/discuss/topic/268299/}} While more concrete than verbal advice, this method floods learners with surface-level differences without indicating which ones are conceptually central or why they matter~\cite{glassman_2015_overcode,moreno_2015_drscratch}.

\para{Rule-based suggestions.} Automated systems such as \texttt{LitterBox}, \texttt{iSnap}, and \texttt{CATNIP} identify predefined Scratch error patterns and generate short, templated hints. These tools succeed for canonical mistakes but collapse on open-ended or novel projects. Their feedback, while precise, remains detached from the learner’s evolving reasoning process~\cite{fraser2021litterbox, fradrich2020commonbugs,boe_2013_hairball, moreno_2015_drscratch, price_2017_isnap_sigcse, marwan_2023_isnap_tlt, fein_2022_catnip}.

\subsection{From Correctness to Comprehension: The Feedback Gap}

Across these diverse forms of help, a consistent shortcoming emerges: current feedback mechanisms optimize for *code correction*, not *conceptual development*. Some forms of support are *overly prescriptive*, performing the cognitive labor for the learner; others are *too abstract*, offering minimal actionable guidance. Both extremes leave students outside the reasoning loop.

We refer to this disconnect as the \emph{feedback gap}: the space between fixing a bug and understanding it. Systems that close this gap must treat debugging not as a terminal outcome but as a structured reasoning process. Feedback should guide the learner through iterative cycles of recognition, explanation, revision, and verification, mirroring how expert instructors teach. In short, the goal is not only to help the program succeed but to help the learner *think*.

\subsection{Toward Step-by-Step, Interactive Feedback}

Bridging the feedback gap requires a shift from static correction to interactive guidance. Rather than delivering a final solution, feedback should engage the learner in an unfolding dialogue: identifying a single issue, explaining its cause, prompting a targeted change, and validating the outcome before proceeding. This iterative approach supports reflection, maintains engagement, and gradually builds transferable debugging skills.

We operationalize this principle in \app, our Step-by-step LLM Guided Debugging system for Scratch. Given a teacher-provided reference project, \app compares a learner’s code against the target, isolates meaningful behavioral differences, and generates contextual explanations using a large language model (Gemini 2.5 Flash Lite). Crucially, \app does not “fix” the code directly. Instead, it produces visual, block-level hints that emphasize *why* each change matters and invites the student to decide *how* to apply it. After each revision, the system re-analyzes the project and continues the dialogue until the learner achieves the intended behavior.

In doing so, \app transforms debugging from a one-time correction into a continuous reasoning experience, narrowing the feedback gap and redefining what effective assistance looks like in block-based programming.

\section{Motivating Examples}
\label{sec:mot}

To validate the prevalence of the feedback gap in authentic learning contexts, we conducted a qualitative review of discussions from the largest public Scratch community forums~\footnote{\url{https://scratch.mit.edu/discuss/}}. We sampled dozens of recent threads in which learners requested debugging help, then selected representative cases that reflect recurring breakdowns in understanding. These examples reveal how well-intentioned help, whether from peers or experts, often produces a correct program but leaves learners without insight into *why* it works. Each illustrates a distinct failure mode of current feedback practice and motivates our design goals for \app.

\para{Example 1: Code without explanation.~\footnote{\url{https://scratch.mit.edu/discuss/topic/805836/}}}
A learner asked how to find the number in a list closest to a given input. A senior member promptly posted a concise, optimized script but declined to explain, replying that they “didn’t have time to go step-by-step.” The code worked, yet the learner immediately followed up: ``Can you tell me why this is faster?'', a question left unanswered. This interaction captures the most common pattern in our review: the fix replaces reasoning. Learners achieve correctness but remain unable to reconstruct the logic that produced it.

\para{Example 2: Correct advice, missing causality.~\footnote{\url{https://scratch.mit.edu/discuss/topic/14843/}}}
A beginner struggled with collision detection; the helper advised, ``Put the check inside a Forever loop.'' The bug disappeared, but the learner's next message was: ``Why does that work?'' Only after a subsequent explanation did understanding emerge, highlighting that continuous checks are required to capture ongoing events. The episode demonstrates that even accurate fixes fail when they omit causal reasoning; the \textbf{why} is as essential as the \textbf{what}.

\para{Example 3: Abstraction without grounding.~\footnote{\url{https://scratch.mit.edu/discuss/topic/80826/}}}
A novice wished to create a quiz game where birds appear one at a time. An expert suggested, “Show one bird at a time,” assuming the learner would infer the control logic. The novice replied, ``What do you mean?'' prompting the helper to elaborate: hide all sprites, pick one randomly, then trigger the next round. Only after this breakdown into discrete steps did the learner succeed. The dialogue shows that feedback must unfold sequentially, high-level guidance alone is too abstract to act on.

\para{Example 4: Explanation beyond comprehension.~\footnote{\url{https://scratch.mit.edu/discuss/topic/196004/}}}
In a thread on a rounding algorithm, an expert offered a mathematically rigorous English explanation. The non-native learner responded, ``I still don't understand even after translating. Should I just memorize the code?'' The expert later rewrote the explanation in simpler language, at which point the learner exclaimed, ``Now I get it!'' This case underscores that feedback must match the learner's cognitive and linguistic level—precision is useless without accessibility.

\para{Summary and Design Implications.}
Across these discussions, the same structural problem reappears: learners receive information that resolves the symptom but not the reason. Fix-only responses shortcut reflection; abstract or overly complex ones overwhelm it. Both disconnect learners from the reasoning loop that drives conceptual growth.  

These empirical patterns directly shaped \app's design. To transform debugging from answer delivery into reasoning support, \app engages learners in an iterative dialogue. It compares a student’s project against a teacher-provided reference, identifies the most meaningful behavioral differences, and uses a large language model to generate block-level explanations of why each change is needed. After each learner revision, \app reevaluates the project and continues the cycle. In doing so, \app operationalizes the key insight from these examples: effective feedback must be stepwise and learner-aligned, guiding students through understanding rather than handing them code.

\section{System Design}
\label{sec:design}
\begin{figure*}[t]
    \centering
    \includegraphics[width=\textwidth]{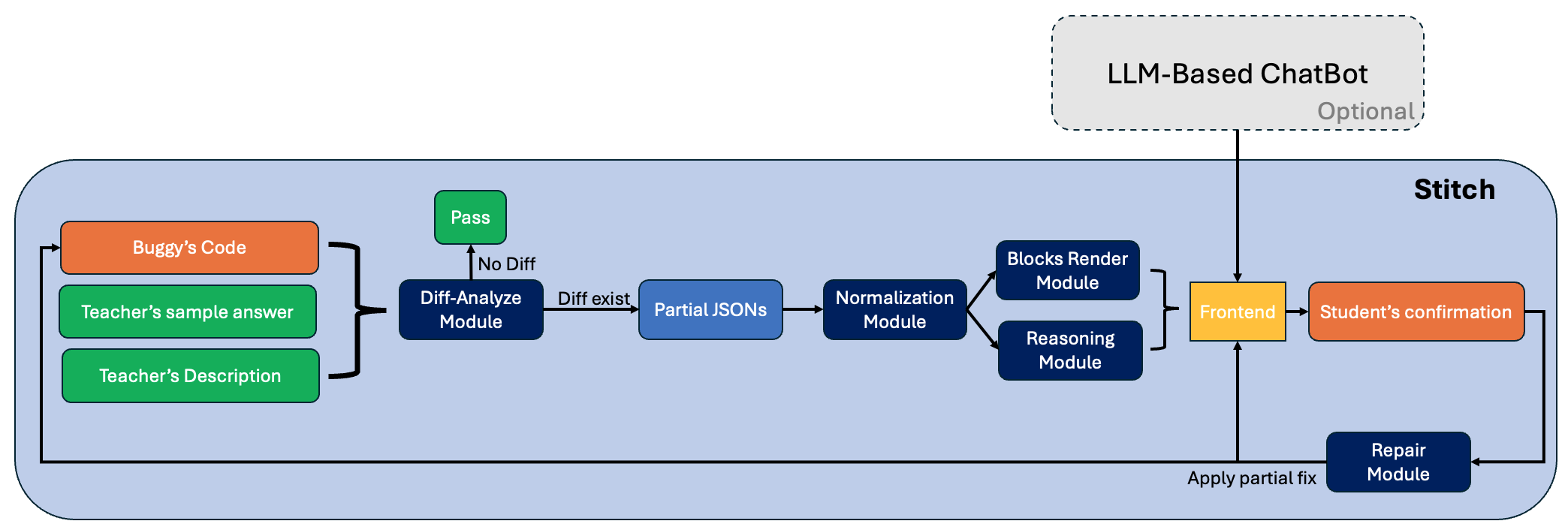}
    \caption{\app pipeline: the system compares student's project and teacher's sample project, identifies differences, generates explanations and block renderings, and guides students iteratively until functional equivalence is achieved.}
    \label{fig:system-flow}
\end{figure*}

Figure~\ref{fig:system-flow} shows the architecture of \app. The system combines a browser extension with a cloud-based backend. Students interact with Scratch as usual, but can request guidance on demand. The extension captures the project code and sends it to the backend, which compares it against a teacher-provided  solution. The backend analyzes differences, generates explanations, and renders visual hints. Results are then displayed in the extension's side panel, where students can either apply suggested fixes or revise their code manually. Afterwards, the updated project is re-analyzed, creating an iterative tutoring loop.  

\label{sec:design}

\subsection{Diff-Analyze Module}

\begin{figure}
    \begin{algorithm}[H]
  \caption{Diff-Analyze Module}
  \begin{algorithmic}[1]
    \Require $S$ (student .sb3), $T$ (teacher .sb3)
    \Ensure $R$ (JSON diff report)
    \State $(A_S,A_T)\gets \mathrm{ParseSB3ToAST}(S,T)$
    \State \Call{Normalize}{$A_S,A_T$} \Comment{noise removal, $\alpha$‑renaming, commutative \& algebraic rewrites}
    \State $M\gets \mathrm{CompareModules}(A_S,A_T)$
           \Comment{stage/sprite roster; flag missing or extra elements}
    \ForAll{sprite $p$ in $\mathrm{MatchSprites}(A_S,A_T)$}
      \State $E\gets \mathrm{MatchScriptsByEvent}(p)$
      \ForAll{matched pair $(s,t)\in E$}
        \State $B\gets \mathrm{AlignBlocksLCS}(s,t)$
               \Comment{LCS‑style on block types/params}
        \State \Call{RecordInsertDeleteModify}{$B$}
      \EndFor
      \ForAll{missing event/script in $E$}
        \State \Call{MarkMissingFunctionality}{}
      \EndFor
    \EndFor
    \State $D\gets \mathrm{Merge}(M,\dots)$ \Comment{Merge all the diffs}
    \State $D\gets \mathrm{FilterSemanticEquivalences}(D)$
    \State $R\gets \mathrm{Normalize}(D)$
    \State \Return $R$
  \end{algorithmic}
\end{algorithm}

\caption{Diff-Analyze Module compares the student’s project with teacher's sample solution, summarizes meaningful differences in sprites, scripts, and blocks, and produces a clear report that drives the feedback shown to the learner.}
    \label{fig:algorithm}
\end{figure}

The Diff-Analyze module compares a student's Scratch project with a teacher's target project and produces a structured summary of differences. Scratch projects are stored in \texttt{.sb3} format, whose core is a JSON file that describes stages, sprites, scripts, and resources. Direct comparison of two JSON files often yields numerous meaningless mismatches (e.g., object IDs, metadata order). To provide meaningful feedback, \app transforms JSON files into an abstract syntax tree (AST) and analyzes differences at multiple semantic levels as described in Figure~\ref{fig:algorithm}. 

\begin{itemize}
    \item \textbf{Module-level.} The system checks macro structures such as stages and character lists. Missing elements (e.g., a sprite in the teacher’s project but absent in the student’s) are flagged, and the student is prompted to add the role or object.  

    \item \textbf{Script-level.} For characters present in both projects, scripts are matched by triggering events (e.g., \texttt{When Green Flag Clicked}). If no corresponding script exists, the student project is marked as missing that functionality.  

     \item \textbf{Block-level.} Within aligned scripts, block sequences are compared while ignoring Scratch-specific IDs. The focus is on block types and parameter behaviors. A sequence alignment algorithm (similar to the Longest Common Subsequence~\footnote{\url{https://www.geeksforgeeks.org/dsa/longest-common-subsequence-dp-4/}}) identifies inserted, deleted, or modified blocks, such as an extra \texttt{Play Sound} or a mismatched conditional operator.  

    \item \textbf{Semantic-level.} To reduce false positives, the module applies normalization (token-trivia removal, $\alpha$-renaming, commutative reordering, algebraic rewrites) before AST differencing. Differences are filtered with rules inspired by refactoring and clone research~\cite{Opdyke1992RefactoringFrameworks,Mens2004SurveyRefactoring,RoyCordy2007CloneSurvey,Falleri2014GumTree,OpenLogicAlphaConversion}, tolerating identifier renaming, safe statement reordering, De Morgan rewrites, and Type-1/2/3 clone variants. When equivalence is uncertain, the system conservatively provides no suggestion.

\end{itemize}

The Diff-Analyze module outputs a JSON report that lists differences at each level, with partial snippets and natural-language explanations (e.g., ``The character \texttt{Cat} is missing the script \texttt{When Green Flag Clicked}''). This report drives subsequent prompts. We assume that the teacher’s target project is correct and complete; thus, the analysis focuses only on student deficiencies. This setup mirrors classroom practices where reference solutions guide step-by-step learning.

\subsection{Normalization Module}
Since Scratch's internal renderer cannot easily handle partial JSON fragments, \app includes a normalization engine. This module translates JSON fragments into structured natural language and maps each opcode to its appearance and parameters. For example, a JSON fragment like \texttt{motion\_movesteps} with input ``3'' is normalized into ``move (3) steps.'' This normalized representation is critical both for rendering blocks visually and for generating prompts that an LLM can understand.  

\subsection{Blocks Rendering Module}
Building on normalization, the rendering module produces graphical block images. Each opcode is mapped to a template styled consistently with Scratch (color, shape, and slot layout). Parameters are filled in, and blocks are stitched vertically to reconstruct script snippets. For instance, the difference between ``move (3) steps'' and ``move (10) steps'' is displayed visually with highlighted parameter values. When rendering fails due to unusual JSON or custom extensions, the system falls back to natural language descriptions. This visual feedback aligns with Scratch’s block-based nature and lowers the barrier for beginners to interpret program differences.  

\subsection{Reasoning Module}
Highlighting differences alone is insufficient; students must understand why these changes matter. The reasoning module leverages a large language model (LLM) to restate differences in plain language and explain their pedagogical significance. For example, instead of merely noting that a “When red flag gets clicked” script is missing, the module clarifies: “Without this script, the sprite never starts moving when the game begins.” By focusing on functional reasoning, the module encourages reflection and understanding, rather than promoting blind replication.

\para{Prompt Engineering.}
During the reasoning stage, the LLM is framed as a patient Scratch instructor with a supportive, non-authoritative tone (e.g., “You are a Scratch programming teacher helping students improve their projects”). Each prompt includes the target code, optional teacher notes, student code, and the specific discrepancy detected. The LLM is instructed to describe the difference, explain why the change is beneficial, and suggest the fix in fewer than 30 words. The style is explicit, yet encouraging, avoiding ambiguity or negativity. Given that Scratch is used by children, the prompts are also filtered for safety and reliability. Feedback is thus positioned as guidance rather than a definitive judgment.
Figure~\ref{fig:prompt_reasoning} illustrates the prompt. For implementation, \app utilizes Google’s Gemini 2.5 Flash Lite~\footnote{\url{https://ai.google.dev/gemini-api/docs/models##gemini-2.5-flash-lite}}, a cost-efficient, high-throughput LLM optimized for interactive tutoring.

\begin{figure}[H]
\centering
\begin{minipage}[t]{\linewidth}
\ttfamily\footnotesize
\textbf{SYSTEM\_Reasoning} \\
\textbf{Role}: Experienced Scratch tutor/evaluator. \\
\textbf{Input}: Teacher JSON, Student JSON, Discrepancy report (Diff), Teacher's description (optional). \\
\textbf{Goal}: Explain the detected difference and why it matters.\\
\textbf{Rules}: Supportive and non-authoritative; avoid negativity; No full code or long edits; name Scratch blocks in quotes (e.g., "When green flag clicked"); Kid-safe and reliable; if uncertain, do not guess.\\
\textbf{Language}: Write in the user's language (fallback English).\\
\textbf{Output}: A concise explanation within 30 words that describe the differences and why it matters. 
\end{minipage}
\caption{Prompt for the reasoning module: generate concise, pedagogically grounded explanations of detected discrepancies.}
\label{fig:prompt_reasoning}
\end{figure}

\subsection{Repair Module}
After receiving feedback, students in \app can choose to revise their code manually or accept suggested fixes. The repair module applies these fixes safely by inserting or replacing partial JSON segments while preserving Scratch's syntax constraints. It also manages the consistency of variable and ID names by scanning the project globally. Once the modified project is repacked and returned, students can see the fix applied immediately, maintaining a smooth learning experience.

\subsection{LLM-based ChatBot}
Some students require more in-depth explanations than what the reasoning module offers. To address this need, \app includes an optional ChatBot feature that allows students to ask questions such as “Why is this change needed?” or “Are there alternative solutions?” The ChatBot is powered by Gemini 2.5 Flash Lite and utilizes longer prompts and curated examples to generate detailed, teacher-like responses. This optional feature empowers students to explore beyond the initial minimal hints, without overwhelming those who prefer brevity.

\para{Prompt Engineering.}
The ChatBot extends the template for richer interaction, with responses that may reach up to 100 words. It is designed to handle follow-up questions such as: “Why is this change needed?”, “Are there alternative solutions?”, or “Can this be generalized for future use?” To support these queries, we drew on real cases from the Scratch forum and crafted examples that guide the LLM to emulate an experienced teacher, linking current projects with broader patterns in programming.
The Prompt Engineering module converts discrepancy reports into LLM prompts, steering the ChatBot to provide pedagogically relevant feedback. High-quality prompt design is crucial for ensuring the effectiveness of this interaction. In \app, we developed templates that combine role definition, contextual grounding, and style constraints, ensuring responses are accurate, supportive, and actionable.
Through this structured prompting process, \app consistently shapes LLM outputs to be educationally sound, safe, and responsive, thus enhancing both student learning outcomes and the overall tutoring experience.
Figure~\ref{fig:prompt_chat} illustrates the prompt. Similarly, \app utilizes Google’s Gemini 2.5 Flash Lite in this module as well.

\begin{figure}[H]
\centering
\begin{minipage}[t]{\linewidth}
\ttfamily\footnotesize

\textbf{SYSTEM\_CHAT} \\
\textbf{Role}: Experienced, patient Scratch tutor/evaluator \\
\textbf{Input}: Student's question, Teacher JSON, Student JSON, Discrepancy report (Diff), Teacher's description (optional). \\
\textbf{Goal}: Answer the student's question; May explain why the change is needed; May offer 1--2 viable alternatives and when to use them; May connect to broader patterns. \\
\textbf{Rules}: Clear and kid-safe; no code dumps or JSON patches; avoid long step-by-step edits; encourage reflection without overwhelming the student; handle follow-up questions gracefully. \\
\textbf{Language}: Write in the user's language (fallback English).\\
\textbf{Output}: A patient and clear text within 100 words that answers student's question.

\end{minipage}
\caption{Prompt for the ChatBot: produce longer, teacher-like responses that clearly answer student's question.}
\label{fig:prompt_chat}
\end{figure}

\subsection{Iterative Tutoring Design.}
The essence of \app is an iterative tutoring loop. Students request guidance, receive feedback, make revisions, and then request guidance again. This process continues until the Diff-Analyze module detects no functional differences. At that point, the system provides summative feedback, such as ``Congratulations, your project now implements all target features.'' By gradually scaffolding corrections, \app ensures students are active participants in debugging rather than passive recipients of solutions.

\subsection{Summary.}
The front-end extension, written in JavaScript, integrates with the Scratch VM to capture the project state and display hints in a side panel, while the backend performs difference analysis, rendering, and prompt generation, invoking Gemini 2.5 Flash Lite for reasoning and dialogue. Asynchronous communication ensures responsiveness: structural differences are typically returned in under half a second, and LLM explanations arrive within a few seconds. Together, these components form a seamless pipeline that analyzes, explains, and guides students through iterative debugging. By combining structured code analysis, visual rendering, LLM-based reasoning, and interactive tutoring loops, \app operationalizes the educational principle of scaffolding at scale in block-based programming education.
\section{Evaluation}
\label{sec:eval}

In this section, we conduct an empirical study on the pedagogical value of \app against another state-of-the-art tool \texttt{iSnap}~\footnote{\url{https://github.com/thomaswp/iSnap}}, an intelligent tutoring system for block-based programming that provides data-driven hints to guide students when they get stuck. We recruit learners to work on real, buggy Scratch projects, and run randomized experiments to measure the tool's performance as well as the enhancement in user's program understanding.

\subsection{Experimental setup}

\para{Experiment participants.}  
We recruit 20 adult participants, who are all college students with some programming experience but no formal training in Scratch. Each participant has completed at least two post-secondary computer science courses, and all have ability to debug code written by popular languages such as \texttt{Python} and \texttt{C}. 
Participants with limited prior exposure to Scratch were intentionally selected to control for prior tool familiarity. This design choice allows us to isolate the pedagogical effectiveness of \textsc{Stitch}'s step-by-step tutoring, ensuring that observed learning gains stem from the tutoring process itself rather than pre-existing knowledge of the Scratch interface.

\para{Buggy Scratch projects selection.} We select 10 real, buggy Scratch Projects from the official Scratch forum~\footnote{\url{https://scratch.mit.edu/explore/projects/all}}. Each project contains exactly one bug that prevents the program from functioning as intended. The bug types include state management errors, conditional logic errors, message mismatch errors, and so on. The programs are reasonably complex, with on average 150 blocks and 7 sprites per project, which makes them challenging to debug. Table~\ref{fig:Dataset Descr.} shows a summary of these projects.

\begin{table}[h]
  \centering
  \begin{tabularx}{\linewidth}{@{} l X l @{}}
    \toprule
    \textbf{Project Name} & \textbf{Bug Description}\\
    \midrule
    Clone Wars & Cloned spites accumulate over time \\
    Bat Maze & Wrong picked color \\
    Ping Pong & Moving scripts missed \\
    Maze Game & Wrong respawn location \\
    Apple Farm & Missing re-transmission logic \\
    Super Mario & Reversed logic condition \\
    Snake Game & Wrong death determination logic \\
    Scratch Clicker & Value carries over after reset \\
    Cat Adventure & Message mismatch \\
    Platformer & Misordered scripts \\
    \bottomrule
  \end{tabularx}
  \caption{List of Scratch projects in the experiment and their bug descriptions.}
  \label{fig:Dataset Descr.}
\end{table}

\para{Experiment procedure.} 
We conduct the empirical study in a randomized experiment. Participants are randomly assigned to one of the two groups, one using \app, another using \texttt{iSnap} as the debugging tool. We selected \texttt{iSnap} as our baseline because it represents a state-of-the-art intelligent tutoring system for block-based programming, offering data-driven, next-step hints that have been adopted in prior studies as a benchmark for evaluating automated feedback effectiveness~\cite{price2019comparison}. Both the \app group and \texttt{iSnap} group have 10 participants. Each participant is randomly assigned 5 buggy Scratch projects for debugging experiments. Each experimental session lasts approximately one hour.

For each Scratch project, participants received two versions: one with the buggy code and one with the correct implementation. In the first stage of the procedure, the participant compares the two versions without using any debugging tool, and tries understand the differences between them. If they successfully find the functional differences, the participant answers the questions: ``Why do we make this change? How would the program behave with and without the change?''. These questions were used solely as metacognitive orientation prompts, not as outcome measures. We did not score or analyze these responses; their purpose was to induce brief self-explanations and focus participants on causal mechanisms before completing the Likert items. Prior work shows that eliciting self-explanations and using pre-question/advance-organizer prompts can prime monitoring and reflection and thereby support comprehension and subsequent learning, which motivates our inclusion of these non-scored prompts~\cite{ausubel1960advanceorganizers,richland2009pretesting,bannert2012prompts, chi1994selfexplanations}. Then the participant answers a set of questions (see following paragraph) to measure their understanding of the error and to gauge their overall debugging experience. In the second stage, we provide the participant with the tool (\app or \texttt{iSnap}) to help them locate and understand the differences between the buggy and the correct code implementations. After using the tool, the participant answers another set of questions (see the following paragraph) to measure their understanding of the error and overall debugging experience with the help of the tool. In each step, they are given up to five minutes to read and compare the programs to maintain experimental consistency across participants. Previous studies on Scratch debugging find meaningful differences in learner debugging performance in time windows of similar length~\cite{greifenstein2021effects}.

\para{Questionnaire and scoring.}
We design two sets of questions for the two experimental stages of each project respectively. Participants respond to each one of the questions on a scale of 5 (1=very poor/strongly disagree; 3=neutral; 5=very good/strongly agree), to indicate to what extent they agree with the statements. The question statements are designed to evaluate participants' understanding of the buggy Scratch projects from different perspectives. 
\begin{itemize}
    \item For each buggy project presented, before using any debugging tool, participants are asked to rate the following statements:
    \begin{itemize}
        \item Q1: Bug Understanding: ``I clearly understand the causes and consequences of the bug.”
        \item Q2: Debugging Confidence: ``I feel confident in locating and fixing the bug.”
        \item Q3: Diff Locating: ``By comparing the buggy and fixed code versions, I can find the key differences in code.”
        \item Q4: Diff Understanding: ``I understand how the differences between the buggy and correct code versions affect program functionality.”
    \end{itemize}
    \item For each buggy project presented, after using the debugging tool (\app or \texttt{iSnap}), participants are asked to rate the following statements:
    \begin{itemize}
        \item Q1: Bug Understanding: ``I clearly understand the causes and consequences of the bug.”
        \item Q2: Debugging Confidence: ``I feel confident in locating and fixing the bug.” 
        \item Q3: Effort Saving: ``Using the tool's hints helped me save time locating the error.”
        \item Q4: Tool Helpfulness: ``I find the hints provided by the debugging tool helpful in resolving the bug.” 
        \item Q5: Hint Clarity: ``The tool's generated feedback is clear and understandable, making it easy for me to understand what to fix.”
    \end{itemize}
\end{itemize}
Among these questions, Q1 and Q2 assess participants' understanding and debugging ability of the specific bugs respectively. The score changes of these two questions after using the tool measure the pedagogical value and satisfaction of learning experience. We use Mann-Whitney U test when comparing scores across experiment groups. In addition, we document any qualitative feedback on the tools from the participants.

\para{Implementation.}
\app was constructed with Python modules and publicly available libraries. The resulting system contains about 1200 lines of code, substantially smaller than existing Scratch feedback frameworks~\cite{deiner2024nuzzlebug,schweikl2025repurr, price2017isnap}. All evaluations were performed with a sampling temperature of 1.0 on a standard Mac mini powered by an Apple M4 processor and 16 GB memory.

\para{Data Availability.}
We provide a package containing a curated dataset of ten Scratch projects, each including the original buggy program and the corresponding teacher-provided sample solution, along with an access link.
All materials are deposited at \url{https://zenodo.org/records/17447624}.  
Due to institutional policy restrictions, the full \app implementation cannot be publicly released at this stage.  
Upon acceptance, we will release the dataset and anonymized user-study results, and will offer controlled access to the system under appropriate research agreements.  
This arrangement ensures reproducibility of our evaluation while complying with institutional requirements.

\subsection{Results}  

\begin{figure}
    \centering
    \includegraphics[page=1,width=\columnwidth]{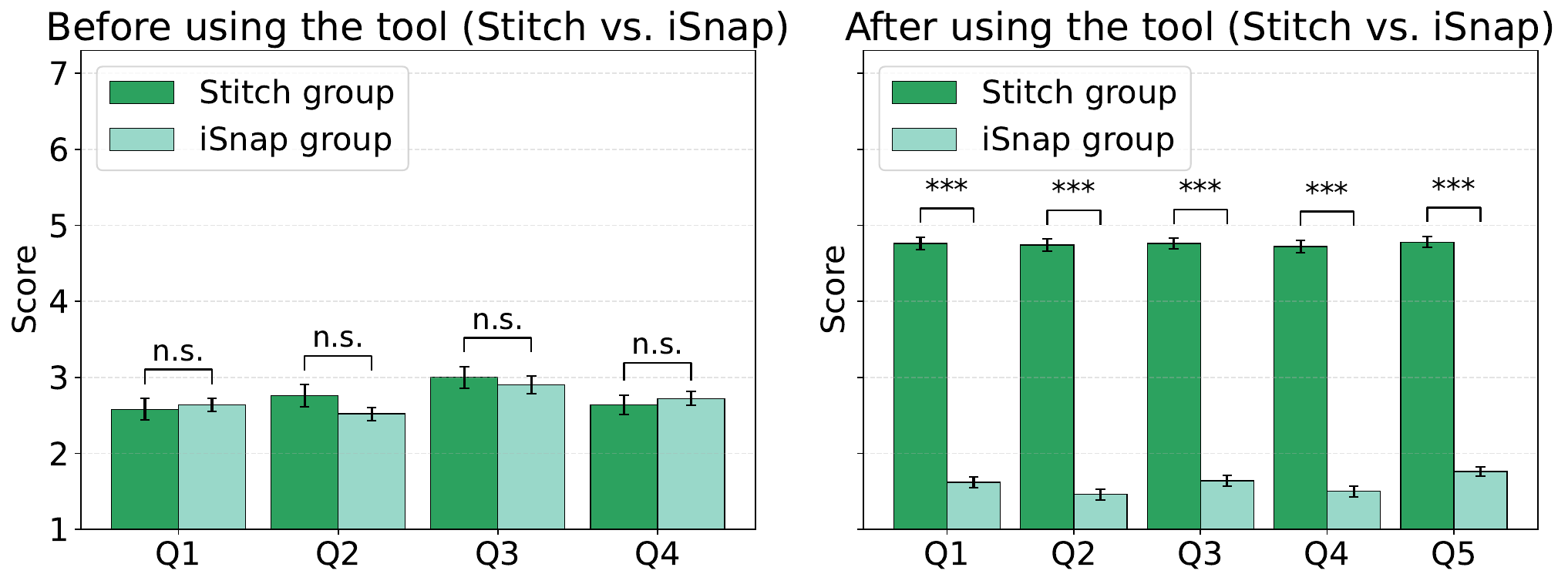}
    \caption{Averaged scores over participants and projects in both the \app and \texttt{iSnap} groups of two sets of different questions measuring participant's understanding and debugging ability of the buggy projects (only Q1 and Q2 are the same), sent before (left) and after (right) using the assigned tutoring tool respectively. The scores are on a scale of 1-5. Before using the tool, the scores are close between the two experiment groups, whereas aftering using the tool, scores in the \app group are significantly higher than those in the \texttt{iSnap} group. Statistical significances are derived with Mann-Whitney U test.}
    \label{fig:stages-results}
\end{figure}

\para{Before using any tutoring system, the average scores to the questions are similar across experiment groups.} Figure \ref{fig:stages-results} left panel shows for each experiment group and each question answered before using the tutoring tool, the bar plots of the average scores over (participant, project) pairs. Participants from both the \app and \texttt{iSnap} groups show relatively low scores in program understanding, debugging confidence, and understanding of the difference between the buggy and correct code versions, without the help of tutoring tool. The average scores to these questions are between 2.5 and 3.0, without significant differences between the \app and \texttt{iSnap} groups ($p{=}0.64$). Therefore, we are able to justify the validity of randomization and the reliability of the scoring mechanism, which sets comparable baselines for studying the effects brought by the tutoring tools in the second step of the experiment sessions.

\para{With the help of the tutoring system, the average scores to the questions are significantly higher in the \app group compared to the \texttt{iSnap} group.} We next compare the results after the participants receive support from the assigned tutoring tool, as shown in Figure \ref{fig:stages-results} right panel. Participants who received guidance from \app on average report significantly higher scores across all the questions compared to those who used \texttt{iSnap}: Q1 Bug understanding: 4.76 in the \app group, 1.62 in the \texttt{iSnap} group, $p=2.04\mathrm{e}{-19}$;  Q2 Debugging confidence: 4.74 in the \app group, 1.46 in the \texttt{iSnap} group, $p=1.30\mathrm{e}{-19}$; Q3 Effort saving: 4.76 in the \app group, 1.64 in the \texttt{iSnap} group, $p=1.15\mathrm{e}{-19}$; Q4 Tool helpfulness: 4.72 in the \app group, 1.50 in the \texttt{iSnap} group, $p=4.03\mathrm{e}{-19}$; Q5 Hint clarity: 4.78 in the \app group, 1.76 in the \texttt{iSnap} group, $p=3.18\mathrm{e}{-20}$. We received positive qualitative feedback from participants in the \app group, e.g. ``\textit{\app is very helpful and in particular, it can help locate the bug with clear explanation, making me much more confident in the debugging process.}''. We conclude that \app outperforms \texttt{iSnap} representing a state-of-the-art tutoring tool from multiple perspectives, including clarity of guidance, learning outcomes, and ease of use.

\begin{figure}
    \centering
    \includegraphics[page=1,width=\columnwidth]{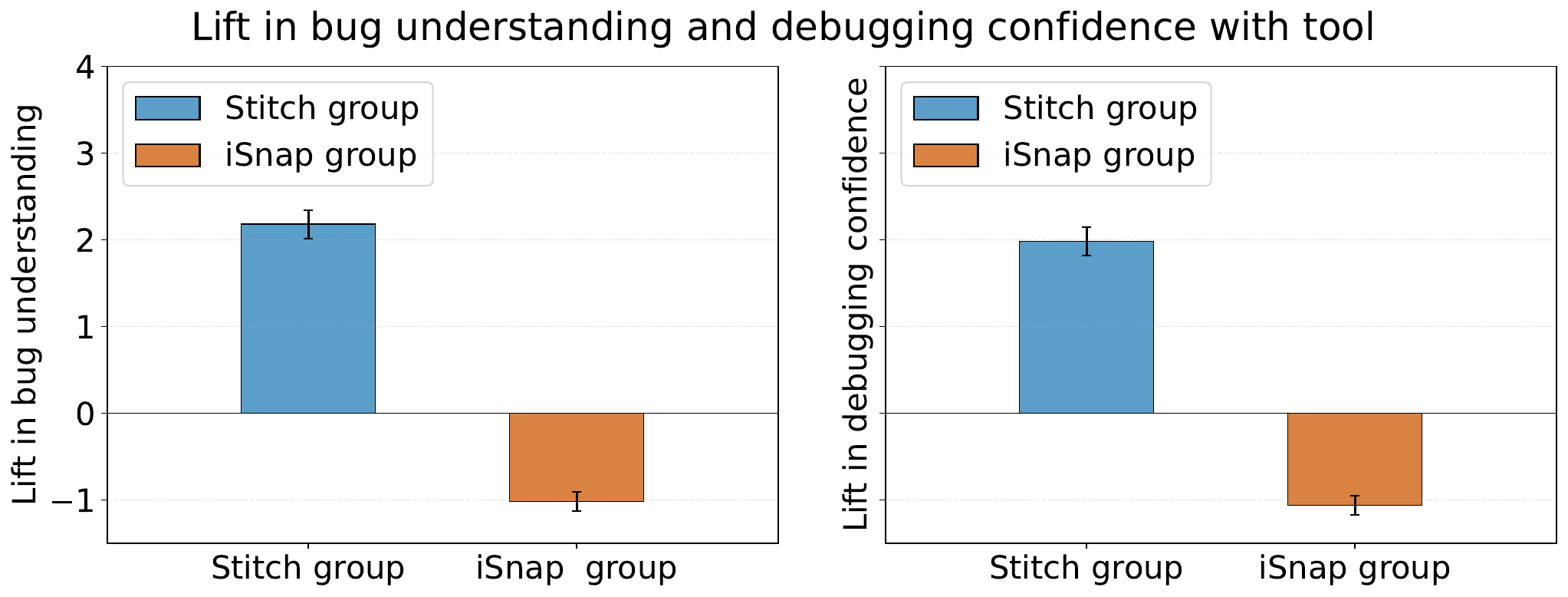}
    \caption{Lift in participants' bug understanding (as measured by the delta in Q1 after using the assigned tutoring tool, left panel), and lift in participants' debugging confidence (as measured by the delta in Q2 after using the assigned tutoring tool, right panel). \app significantly improves participants' bug understanding and debugging confidence, while the \texttt{iSnap} group sees a decline.}
    \label{fig:effect}
\end{figure}

\para{\app significantly enhances participants' understanding and debugging confidence of the buggy projects.} We specifically compare the results of Q1 and Q2 which are asked both before and after using the tutoring tool. For each (participant, project) pair, we calculate the change in the scores of Q1 and Q2 after providing the tutoring tool. The score change in Q1 measures participant's improvement in bug understanding, and the score change in Q2 measures participant's lift in debugging confidence. We compare the differences in these lifts across experiment groups, as shown in Figure~\ref{fig:effect}. For bug understanding, we see an average lift of 2.18 ($p{=}8.68\mathrm{e}-16$) in the \app group, while an average decline of 1.02 ($p{=}3.15\mathrm{e}-12$) in the \texttt{iSnap} group. For debugging confidence, we see an average lift of 1.98 ($p{=}2.87\mathrm{e}-15$) in the \app group, while an average decline of 1.06 ($p{=}2.03\mathrm{e}-12$) in the \texttt{iSnap} group. Based on the qualitative feedback received from participants in the \texttt{iSnap} group, e.g. ``\textit{There is no clear next step and the hints do not align with the actual fix.}'', learners can be easily confused by the guidance provided by \texttt{iSnap} partially due to its direct-answer style of interaction, especially when the buggy project is complex to fix. In contrast, \app's step-by-step tutoring style leads to a remarkable boost in both bug understanding and debugging confidence than the traditional approaches, demonstrating its pedagogical value.

\section{Threats to Validity}

\para{Internal validity.}
Our findings and comparisons may be influenced by specific design choices and the experimental setup. \app’s reliance on a particular LLM (Gemini 2.5 Flash Lite) introduces nondeterminism, i.e., repeated runs may yield slightly different results. We mitigate this by fixing the prompt template and guidance format, but the nondeterministic nature of LLM outputs remains a source of variability. In addition, the example tasks used to demonstrate \app are drawn from forum cases with known correct versions available for comparison. While these examples reflect real-world problems, this selection may inadvertently favor our difference-based approach, potentially overestimating its effectiveness on more arbitrary or unconstrained Scratch projects. 

\para{External validity.}
\app currently assumes that teachers provide a sample code as a reference solution. This assumption aligns with many classroom scenarios (e.g., homework assignments with well-defined solutions), but it does not cover open-ended tasks where only a high-level description is provided and multiple solutions may be valid. Under such circumstances, the difference-based paradigm may be less appropriate. An alternative design would guide students against requirement specifications or test cases rather than a single reference implementation. Extending \app to address more open-ended pedagogical scenarios represents an important direction for future work. Moreover, since the participants have some prior programming experience but little familiarity with Scratch, their behaviors may not fully simulate those of young novice learners. Future research will include beginners and school-age learners to evaluate \app's effectiveness across a broader range of age and experience levels.

\para{Correctness validity.}
Our evaluation measures learning outcomes based on the successful completion of projects and subjective feedback from human participants. While these measures provide evidence of pedagogical value, they may not fully capture long-term learning outcomes or transferability of debugging skills. Moreover, our assessment of ``correctness'' relies on matching functionality to the reference project, which assumes that the reference indeed produces the intended behavior. To ensure this, we validated projects through execution in the Scratch VM and reviewed the outcomes manually. Nonetheless, alternative correct solutions could exist, and some hidden bugs may not have been detected. Future studies with larger and more diverse learner populations are needed to further justify the validity of these claims.

\section{Related Work}

\para{AI Copilots.} Building on the success of AI coding assistants in text-based programming (e.g., GitHub Copilot), recent systems adapt copilots to block-based environments~\cite{griebl2023blockllm}. ChatScratch~\cite{chen2024chatscratch} embeds LLMs into storyboarding tools to help children design games and generate sprites or backdrops from descriptions, while MindScratch~\cite{chen2025mindscratch} supports classroom mind mapping linked to learning goals. Interactive debuggers such as NuzzleBug~\cite{deiner2024nuzzlebug} and Blink~\cite{STRIJBOL2024101617} add stepping, pausing, and reverse execution for clearer behavior visualization. RePurr~\cite{schweikl2025repurr} performs automated repair via genetic programming guided by fault localization.
Static analyzers detect novice issues (dead code, missing handlers, unused broadcasts)~\cite{boe2013hairball,techapalokul2017qualityhound,moreno2015drscratch,fradrich2020commonbugs,fraser2021litterbox, obermueller2024fixes,shao2025llmscorrectlyintegratedsoftware}, and testing frameworks generate inputs or assertions to expose behavioral errors~\cite{johnson2016itch,stahlbauer2019whisker,deiner2023autotest,goetz2022modelbased,wang2021snapcheck,obermuller2021catnip,zhang2025systematicstudytimelimit}. Cognimates Scratch Copilot~\cite{druga2025scratch} embeds an AI assistant within Scratch to provide real-time, conversational guidance during idea generation and code creation, with a focus on open-ended creativity rather than goal-directed repair.
Intelligent tutors for text-based languages typically use templates or rule-based progressions from syntax to logic errors. In contrast, \app couples LLM reasoning with differential analysis to generate adaptive, context-aware feedback precisely aligned with task objectives, achieving a level of intelligent guidance absent in prior programming assistants.

\para{Automated Evaluation and Feedback Tools.}
Automated code~\cite{Santolucito22} and feedback have long been central to programming education, spanning online grading systems~\cite{zhang2025systematicstudytimelimit}, unit-test-based feedback~\cite{gmerge} for text programming, and specialized assessment tools for visual environments like Scratch. Dr.~Scratch~\cite{moreno2015drscratch}, for instance, performs static analysis to assess computational thinking by examining block types and control structures, offering high-level suggestions for improvement. However, its feedback remains coarse-grained and detached from specific project goals~\cite{obermuller2023effects}.
Other tools detect inefficiencies or misconfigurations~\cite{ConfigX} such as unused variables or inactive event blocks, offering static warnings that improve code quality but lack interactivity~\cite{caspari2023scratchlog}. \app advances this approach with conversational, multi-step feedback that adapts to student progress. Similar to Intelligent Tutoring Systems (ITS), it delivers adaptive guidance, but unlike rule-based ITS, \app uses AI-generated prompts for flexible, goal-aware instruction.

\para{Guided Learning and Scaffolding Theory.}
This work builds on educational theories of scaffolding and constructivist learning~\cite{WoodBrunerRoss1976}, which emphasize guided exploration rather than direct instruction. In Scratch learning, teachers are encouraged to support students in transforming ideas into projects while gradually releasing control to foster confidence~\cite{druga2025scratch,greifenstein2021hints}. \app is designed to emulate this scaffolding process: it provides timely help when learners face obstacles but reduces intervention as they gain mastery. The interactive design instantiates this principle: \app reveals differences step by step instead of offering full explanations at once~\cite{greifenstein2023hintcontent}. We also draw on research in child–computer interaction to shape the assistant’s tone and behavior, ensuring responses remain patient, encouraging, and respectful of young learners’ autonomy.

\para{Automated Feedback in Educational Programming.}
Automated feedback systems~\cite{ko2004whyline,barr2014plastic,keuning2019automatedfeedback,liffiton2023codehelp,clef,PyDex,si2025viscratchusinglargelanguage} for novice programmers often rely on rule-based hints derived from expert-authored solutions. In block-based environments, iSnap~\cite{price2017isnap} and ITAP~\cite{rivers2017itap} compare student's progress against expert solutions, while CATNIP~\cite{fein2022catnip} generates Scratch hints using instructor-defined test suites to detect behavioral mismatches.
Recent work~\cite{gmerge} has explored applying LLMs to question answering, test analysis, and conversational practice, but children’s programming remains underexplored. Scratch Copilot~\cite{druga2025scratch} pioneers LLM-based multimodal feedback for Scratch. Our work extends this direction, showing that LLMs can generate not only code or repairs but also pedagogically structured dialogues that function as inspirational tutors. To ensure reliability, \app combines LLM reasoning with differential program analysis, constraining the model’s focus to reducing errors while retaining natural, human-like communication.

\section{Conclusion}

We present \app, a Scratch tutoring system that integrates difference analysis with LLMs to deliver step-by-step, conversational feedback. By aligning learner's code with a teacher's sample answer, \app combines the “diff” concept from software engineering with the pedagogical principle of ``scaffolding'', to enable a novel form of AI-powered tutoring in visual programming. In particular, \app emphasizes learner's engagement and skill development, instead of a single-step bug fixing.
Our results highlight the potential of LLMs to act not only as the behind-the-scene code fixers but as interactive mentors that guide learners through their problem-solving journey. We envision extensions of this approach to other block-based environments and broader science learning, where AI decision-making can complement teachers' instructional strategies. By prioritizing learning experience over automation, \app marks an important step towards the next generation of intelligent tutoring systems and tools that enhance teaching efficiency while fostering deeper learner engagement.

\newpage



\bibliographystyle{ACM-Reference-Format}
\bibliography{scratch}


\begin{thebibliography}{65}


\ifx \showCODEN    \undefined \def \showCODEN     #1{\unskip}     \fi
\ifx \showISBNx    \undefined \def \showISBNx     #1{\unskip}     \fi
\ifx \showISBNxiii \undefined \def \showISBNxiii  #1{\unskip}     \fi
\ifx \showISSN     \undefined \def \showISSN      #1{\unskip}     \fi
\ifx \showLCCN     \undefined \def \showLCCN      #1{\unskip}     \fi
\ifx \shownote     \undefined \def \shownote      #1{#1}          \fi
\ifx \showarticletitle \undefined \def \showarticletitle #1{#1}   \fi
\ifx \showURL      \undefined \def \showURL       {\relax}        \fi
\providecommand\bibfield[2]{#2}
\providecommand\bibinfo[2]{#2}
\providecommand\natexlab[1]{#1}
\providecommand\showeprint[2][]{arXiv:#2}

\bibitem[Aleven et~al\mbox{.}(2016)]%
        {aleven_2016_help_seeking}
\bibfield{author}{\bibinfo{person}{Vincent Aleven}, \bibinfo{person}{Ido Roll}, \bibinfo{person}{Bruce~M. McLaren}, {and} \bibinfo{person}{Kenneth~R. Koedinger}.} \bibinfo{year}{2016}\natexlab{}.
\newblock \showarticletitle{Help Helps, But Only So Much: Research on Help Seeking with Intelligent Tutoring Systems}.
\newblock \bibinfo{journal}{\emph{International Journal of Artificial Intelligence in Education}} \bibinfo{volume}{26}, \bibinfo{number}{1} (\bibinfo{year}{2016}), \bibinfo{pages}{205--250}.
\newblock
\href{https://doi.org/10.1007/s40593-015-0089-1}{doi:\nolinkurl{10.1007/s40593-015-0089-1}}


\bibitem[Ausubel(1960)]%
        {ausubel1960advanceorganizers}
\bibfield{author}{\bibinfo{person}{David~P. Ausubel}.} \bibinfo{year}{1960}\natexlab{}.
\newblock \showarticletitle{The Use of Advance Organizers in the Learning and Retention of Meaningful Verbal Material}.
\newblock \bibinfo{journal}{\emph{Journal of Educational Psychology}} \bibinfo{volume}{51}, \bibinfo{number}{5} (\bibinfo{year}{1960}), \bibinfo{pages}{267--272}.
\newblock
\href{https://doi.org/10.1037/h0046669}{doi:\nolinkurl{10.1037/h0046669}}


\bibitem[Bannert and Reimann(2012)]%
        {bannert2012prompts}
\bibfield{author}{\bibinfo{person}{Maria Bannert} {and} \bibinfo{person}{Peter Reimann}.} \bibinfo{year}{2012}\natexlab{}.
\newblock \showarticletitle{Supporting self-regulated hypermedia learning through prompts}.
\newblock \bibinfo{journal}{\emph{Instructional Science}} \bibinfo{volume}{40}, \bibinfo{number}{1} (\bibinfo{year}{2012}), \bibinfo{pages}{193--211}.
\newblock
\href{https://doi.org/10.1007/s11251-011-9167-4}{doi:\nolinkurl{10.1007/s11251-011-9167-4}}


\bibitem[Barr et~al\mbox{.}(2014)]%
        {barr2014plastic}
\bibfield{author}{\bibinfo{person}{Earl~T. Barr}, \bibinfo{person}{Yuriy Brun}, \bibinfo{person}{Premkumar~T. Devanbu}, \bibinfo{person}{Mark Harman}, {and} \bibinfo{person}{Federica Sarro}.} \bibinfo{year}{2014}\natexlab{}.
\newblock \showarticletitle{The Plastic Surgery Hypothesis}. In \bibinfo{booktitle}{\emph{FSE}}. \bibinfo{pages}{306--317}.
\newblock
\href{https://doi.org/10.1145/2635868.2635898}{doi:\nolinkurl{10.1145/2635868.2635898}}


\bibitem[Boe et~al\mbox{.}(2013a)]%
        {boe_2013_hairball}
\bibfield{author}{\bibinfo{person}{Bryce Boe}, \bibinfo{person}{Charlotte Hill}, \bibinfo{person}{Michelle Len}, \bibinfo{person}{Greg Dreschler}, \bibinfo{person}{Phillip Conrad}, {and} \bibinfo{person}{Diana Franklin}.} \bibinfo{year}{2013}\natexlab{a}.
\newblock \showarticletitle{Hairball: Lint-inspired Static Analysis of Scratch Projects}. In \bibinfo{booktitle}{\emph{Proceedings of SIGCSE 2013}}. \bibinfo{pages}{215--220}.
\newblock
\href{https://doi.org/10.1145/2445196.2445265}{doi:\nolinkurl{10.1145/2445196.2445265}}


\bibitem[Boe et~al\mbox{.}(2013b)]%
        {boe2013hairball}
\bibfield{author}{\bibinfo{person}{Brennan Boe}, \bibinfo{person}{Caitlin Hill}, \bibinfo{person}{Michelle Len}, \bibinfo{person}{Gina Dreschler}, \bibinfo{person}{Philip Conrad}, {and} \bibinfo{person}{Diana Franklin}.} \bibinfo{year}{2013}\natexlab{b}.
\newblock \showarticletitle{Hairball: Lint-Inspired Static Analysis of Scratch Projects}. In \bibinfo{booktitle}{\emph{SIGCSE}}. \bibinfo{pages}{215--220}.
\newblock
\href{https://doi.org/10.1145/2445196.2445265}{doi:\nolinkurl{10.1145/2445196.2445265}}


\bibitem[Caspari et~al\mbox{.}(2023)]%
        {caspari2023scratchlog}
\bibfield{author}{\bibinfo{person}{Laura Caspari}, \bibinfo{person}{Luisa Greifenstein}, \bibinfo{person}{Ute Heuer}, {and} \bibinfo{person}{Gordon Fraser}.} \bibinfo{year}{2023}\natexlab{}.
\newblock \showarticletitle{ScratchLog: Live Learning Analytics for Scratch}. In \bibinfo{booktitle}{\emph{Proceedings of the 2023 ACM Conference on Innovation and Technology in Computer Science Education (ITiCSE '23)}}. \bibinfo{publisher}{ACM}.
\newblock
\href{https://doi.org/10.1145/3587102.3588836}{doi:\nolinkurl{10.1145/3587102.3588836}}


\bibitem[Chen et~al\mbox{.}(2024)]%
        {chen2024chatscratch}
\bibfield{author}{\bibinfo{person}{Liuqing Chen}, \bibinfo{person}{Shuhong Xiao}, \bibinfo{person}{Yunnong Chen}, \bibinfo{person}{Yaxuan Song}, \bibinfo{person}{Ruoyu Wu}, {and} \bibinfo{person}{Lingyun Sun}.} \bibinfo{year}{2024}\natexlab{}.
\newblock \showarticletitle{ChatScratch: An AI-augmented system toward autonomous visual programming learning for children aged 6-12}. In \bibinfo{booktitle}{\emph{Proceedings of the 2024 CHI Conference on Human Factors in Computing Systems}}. \bibinfo{pages}{1--19}.
\newblock


\bibitem[Chen et~al\mbox{.}(2025)]%
        {chen2025mindscratch}
\bibfield{author}{\bibinfo{person}{Yunnong Chen}, \bibinfo{person}{Shuhong Xiao}, \bibinfo{person}{Yaxuan Song}, \bibinfo{person}{Zejian Li}, \bibinfo{person}{Lingyun Sun}, {and} \bibinfo{person}{Liuqing Chen}.} \bibinfo{year}{2025}\natexlab{}.
\newblock \showarticletitle{MindScratch: A Visual Programming Support Tool for Classroom Learning Based on Multimodal Generative AI}.
\newblock \bibinfo{journal}{\emph{International Journal of Human--Computer Interaction}} (\bibinfo{year}{2025}), \bibinfo{pages}{1--19}.
\newblock


\bibitem[Chi et~al\mbox{.}(1994a)]%
        {chi_1994_self_explanations}
\bibfield{author}{\bibinfo{person}{Michelene T.~H. Chi}, \bibinfo{person}{Nicholas De~Leeuw}, \bibinfo{person}{Mei-Hung Chiu}, {and} \bibinfo{person}{Christian LaVancher}.} \bibinfo{year}{1994}\natexlab{a}.
\newblock \showarticletitle{Eliciting Self-Explanations Improves Understanding}.
\newblock \bibinfo{journal}{\emph{Cognitive Science}} \bibinfo{volume}{18}, \bibinfo{number}{3} (\bibinfo{year}{1994}), \bibinfo{pages}{439--477}.
\newblock
\href{https://doi.org/10.1207/s15516709cog1803_3}{doi:\nolinkurl{10.1207/s15516709cog1803_3}}


\bibitem[Chi et~al\mbox{.}(1994b)]%
        {chi1994selfexplanations}
\bibfield{author}{\bibinfo{person}{Michelene T.~H. Chi}, \bibinfo{person}{Nicholas de Leeuw}, \bibinfo{person}{Mei-Hung Chiu}, {and} \bibinfo{person}{Christian LaVancher}.} \bibinfo{year}{1994}\natexlab{b}.
\newblock \showarticletitle{Eliciting Self-Explanations Improves Understanding}.
\newblock \bibinfo{journal}{\emph{Cognitive Science}} \bibinfo{volume}{18}, \bibinfo{number}{3} (\bibinfo{year}{1994}), \bibinfo{pages}{439--477}.
\newblock
\href{https://doi.org/10.1207/s15516709cog1803_3}{doi:\nolinkurl{10.1207/s15516709cog1803_3}}


\bibitem[Deiner et~al\mbox{.}(2023)]%
        {deiner2023autotest}
\bibfield{author}{\bibinfo{person}{Adina Deiner}, \bibinfo{person}{Patric Feldmeier}, \bibinfo{person}{Gordon Fraser}, \bibinfo{person}{Sebastian Schweikl}, {and} \bibinfo{person}{Wengran Wang}.} \bibinfo{year}{2023}\natexlab{}.
\newblock \showarticletitle{Automated Test Generation for Scratch Programs}.
\newblock \bibinfo{journal}{\emph{Empirical Software Engineering}} \bibinfo{volume}{28}, \bibinfo{number}{79} (\bibinfo{year}{2023}).
\newblock
\href{https://doi.org/10.1007/s10664-022-10255-x}{doi:\nolinkurl{10.1007/s10664-022-10255-x}}


\bibitem[Deiner and Fraser(2024)]%
        {deiner2024nuzzlebug}
\bibfield{author}{\bibinfo{person}{Adina Deiner} {and} \bibinfo{person}{Gordon Fraser}.} \bibinfo{year}{2024}\natexlab{}.
\newblock \showarticletitle{NuzzleBug: Debugging Block-Based Programs in Scratch}. In \bibinfo{booktitle}{\emph{Proceedings of the 46th IEEE/ACM International Conference on Software Engineering (ICSE ’24)}}. \bibinfo{pages}{1—2}.
\newblock
\href{https://doi.org/10.1145/3597503.3623331}{doi:\nolinkurl{10.1145/3597503.3623331}}


\bibitem[Druga and Ko(2025)]%
        {druga2025scratch}
\bibfield{author}{\bibinfo{person}{Stefania Druga} {and} \bibinfo{person}{Amy~J Ko}.} \bibinfo{year}{2025}\natexlab{}.
\newblock \showarticletitle{Scratch Copilot: Supporting Youth Creative Coding with AI}.
\newblock In \bibinfo{booktitle}{\emph{Proceedings of the 24th Interaction Design and Children}}. \bibinfo{pages}{140--153}.
\newblock


\bibitem[Elliott et~al\mbox{.}(2023)]%
        {Elliott2023DebuggingPedagogy}
\bibfield{author}{\bibinfo{person}{Colin~Hennessy Elliott}, \bibinfo{person}{Alexandra Gendreau~Chakarov}, \bibinfo{person}{Jeffrey~B. Bush}, \bibinfo{person}{Jessie Nixon}, {and} \bibinfo{person}{Mimi Recker}.} \bibinfo{year}{2023}\natexlab{}.
\newblock \showarticletitle{Toward a debugging pedagogy: helping students learn to get unstuck with physical computing systems}.
\newblock \bibinfo{journal}{\emph{Information and Learning Sciences}} \bibinfo{volume}{124}, \bibinfo{number}{1-2} (\bibinfo{year}{2023}), \bibinfo{pages}{1--24}.
\newblock
\href{https://doi.org/10.1108/ILS-03-2022-0051}{doi:\nolinkurl{10.1108/ILS-03-2022-0051}}


\bibitem[Elliott et~al\mbox{.}(2024)]%
        {Elliott2024DebuggingPedagogies}
\bibfield{author}{\bibinfo{person}{Colin~Hennessy Elliott}, \bibinfo{person}{Jessie Nixon}, \bibinfo{person}{Alexandra Gendreau~Chakarov}, \bibinfo{person}{Jeffrey~B. Bush}, \bibinfo{person}{Michael~J. Schneider}, {and} \bibinfo{person}{Mimi Recker}.} \bibinfo{year}{2024}\natexlab{}.
\newblock \showarticletitle{Characterizing Teacher Support of Debugging with Physical Computing: Debugging Pedagogies in Practice}.
\newblock \bibinfo{journal}{\emph{ACM Transactions on Computing Education}} \bibinfo{volume}{24}, \bibinfo{number}{4} (\bibinfo{year}{2024}), \bibinfo{pages}{Article 48}.
\newblock
\href{https://doi.org/10.1145/3677612}{doi:\nolinkurl{10.1145/3677612}}


\bibitem[Falleri et~al\mbox{.}(2014)]%
        {Falleri2014GumTree}
\bibfield{author}{\bibinfo{person}{Jean-R{\'e}my Falleri}, \bibinfo{person}{Flor{\'e}al Morandat}, \bibinfo{person}{Xavier Blanc}, \bibinfo{person}{Matias Martinez}, {and} \bibinfo{person}{Martin Monperrus}.} \bibinfo{year}{2014}\natexlab{}.
\newblock \showarticletitle{Fine-grained and Accurate Source Code Differencing}. In \bibinfo{booktitle}{\emph{Proceedings of the 29th IEEE/ACM International Conference on Automated Software Engineering (ASE '14)}}. \bibinfo{publisher}{ACM}, \bibinfo{address}{V{\"a}ster{\aa}s, Sweden}, \bibinfo{pages}{313--324}.
\newblock
\href{https://doi.org/10.1145/2642937.2642982}{doi:\nolinkurl{10.1145/2642937.2642982}}


\bibitem[Fein et~al\mbox{.}(2022a)]%
        {fein2022catnip}
\bibfield{author}{\bibinfo{person}{Benedikt Fein}, \bibinfo{person}{Florian Oberm{\"u}ller}, {and} \bibinfo{person}{Gordon Fraser}.} \bibinfo{year}{2022}\natexlab{a}.
\newblock \showarticletitle{CATNIP: An Automated Hint Generation Tool for Scratch}. In \bibinfo{booktitle}{\emph{ITiCSE}}. \bibinfo{pages}{124--130}.
\newblock


\bibitem[Fein et~al\mbox{.}(2022b)]%
        {fein_2022_catnip}
\bibfield{author}{\bibinfo{person}{Benedikt Fein}, \bibinfo{person}{Florian Oberm{\"u}ller}, {and} \bibinfo{person}{Gordon Fraser}.} \bibinfo{year}{2022}\natexlab{b}.
\newblock \showarticletitle{CATNIP: An Automated Hint Generation Tool for Scratch}. In \bibinfo{booktitle}{\emph{Proceedings of ITiCSE 2022}}.
\newblock
\href{https://doi.org/10.1145/3502718.3524820}{doi:\nolinkurl{10.1145/3502718.3524820}}


\bibitem[Fr{\"a}drich et~al\mbox{.}(2020)]%
        {fradrich2020commonbugs}
\bibfield{author}{\bibinfo{person}{Christoph Fr{\"a}drich}, \bibinfo{person}{Florian Oberm{\"u}ller}, \bibinfo{person}{Nina K{\"o}rber}, \bibinfo{person}{Ute Heuer}, {and} \bibinfo{person}{Gordon Fraser}.} \bibinfo{year}{2020}\natexlab{}.
\newblock \showarticletitle{Common Bugs in Scratch Programs}. In \bibinfo{booktitle}{\emph{ITiCSE}}. \bibinfo{pages}{89--95}.
\newblock
\href{https://doi.org/10.1145/3341525.3387389}{doi:\nolinkurl{10.1145/3341525.3387389}}


\bibitem[Fraser et~al\mbox{.}(2021)]%
        {fraser2021litterbox}
\bibfield{author}{\bibinfo{person}{Gordon Fraser}, \bibinfo{person}{Ute Heuer}, \bibinfo{person}{Nina K{\"o}rber}, \bibinfo{person}{Florian Oberm{\"u}ller}, {and} \bibinfo{person}{Ewald Wasmeier}.} \bibinfo{year}{2021}\natexlab{}.
\newblock \showarticletitle{LitterBox: A Linter for Scratch Programs}.
\newblock \bibinfo{journal}{\emph{Proceedings of the 2021 IEEE/ACM 43rd International Conference on Software Engineering: Software Engineering Education and Training (ICSE-SEET)}} (\bibinfo{year}{2021}), \bibinfo{pages}{182--188}.
\newblock
\href{https://doi.org/10.1109/ICSE-SEET52601.2021.00028}{doi:\nolinkurl{10.1109/ICSE-SEET52601.2021.00028}}


\bibitem[Glassman et~al\mbox{.}(2015)]%
        {glassman_2015_overcode}
\bibfield{author}{\bibinfo{person}{Elena~L. Glassman}, \bibinfo{person}{Jeremy Scott}, \bibinfo{person}{Rishabh Singh}, \bibinfo{person}{Philip~J. Guo}, {and} \bibinfo{person}{Robert~C. Miller}.} \bibinfo{year}{2015}\natexlab{}.
\newblock \showarticletitle{OverCode: Visualizing Variation in Student Solutions to Programming Problems at Scale}.
\newblock \bibinfo{journal}{\emph{ACM Transactions on Computer-Human Interaction (TOCHI)}} \bibinfo{volume}{22}, \bibinfo{number}{2} (\bibinfo{year}{2015}).
\newblock
\href{https://doi.org/10.1145/2699751}{doi:\nolinkurl{10.1145/2699751}}


\bibitem[G{\"o}tz et~al\mbox{.}(2022)]%
        {goetz2022modelbased}
\bibfield{author}{\bibinfo{person}{Katharina G{\"o}tz}, \bibinfo{person}{Patric Feldmeier}, {and} \bibinfo{person}{Gordon Fraser}.} \bibinfo{year}{2022}\natexlab{}.
\newblock \showarticletitle{Model-based Testing of Scratch Programs}. In \bibinfo{booktitle}{\emph{IEEE ICST}}. \bibinfo{pages}{411--421}.
\newblock


\bibitem[Greifenstein et~al\mbox{.}(2023)]%
        {greifenstein2023hintcontent}
\bibfield{author}{\bibinfo{person}{Luisa Greifenstein}, \bibinfo{person}{Markus Brune}, \bibinfo{person}{Tobias Fuchs}, \bibinfo{person}{Ute Heuer}, {and} \bibinfo{person}{Gordon Fraser}.} \bibinfo{year}{2023}\natexlab{}.
\newblock \showarticletitle{Impact of Hint Content on Performance and Learning: A Study with Primary School Children in a Scratch Course}. In \bibinfo{booktitle}{\emph{Proceedings of the 18th WiPSCE Conference on Primary and Secondary Computing Education Research (WiPSCE '23)}}. \bibinfo{publisher}{ACM}, \bibinfo{pages}{1--10}.
\newblock


\bibitem[Greifenstein et~al\mbox{.}(2021a)]%
        {greifenstein2021effects}
\bibfield{author}{\bibinfo{person}{Luisa Greifenstein}, \bibinfo{person}{Florian Oberm{\"u}ller}, \bibinfo{person}{Ewald Wasmeier}, \bibinfo{person}{Ute Heuer}, {and} \bibinfo{person}{Gordon Fraser}.} \bibinfo{year}{2021}\natexlab{a}.
\newblock \showarticletitle{Effects of hints on debugging Scratch programs: An empirical study with primary school teachers in training}. In \bibinfo{booktitle}{\emph{Proceedings of the 16th Workshop in Primary and Secondary Computing Education}}. \bibinfo{pages}{1--10}.
\newblock


\bibitem[Greifenstein et~al\mbox{.}(2021b)]%
        {greifenstein2021hints}
\bibfield{author}{\bibinfo{person}{Luisa Greifenstein}, \bibinfo{person}{Florian Oberm{\"u}ller}, \bibinfo{person}{Ewald Wasmeier}, \bibinfo{person}{Ute Heuer}, {and} \bibinfo{person}{Gordon Fraser}.} \bibinfo{year}{2021}\natexlab{b}.
\newblock \showarticletitle{Effects of Hints on Debugging Scratch Programs: An Empirical Study with Primary School Teachers in Training}. In \bibinfo{booktitle}{\emph{Proceedings of the 16th Workshop in Primary and Secondary Computing Education (WiPSCE '21)}}. \bibinfo{publisher}{ACM}.
\newblock
\href{https://doi.org/10.1145/3481312.3481344}{doi:\nolinkurl{10.1145/3481312.3481344}}


\bibitem[Griebl et~al\mbox{.}(2023)]%
        {griebl2023blockllm}
\bibfield{author}{\bibinfo{person}{Erik Griebl}, \bibinfo{person}{Benedikt~S. Clegg}, \bibinfo{person}{Florian Oberm{\"u}ller}, \bibinfo{person}{Gordon Fraser}, \bibinfo{person}{René Just}, {and} \bibinfo{person}{Phil McMinn}.} \bibinfo{year}{2023}\natexlab{}.
\newblock \showarticletitle{On the Applicability of Language Models to Block-Based Programs}. In \bibinfo{booktitle}{\emph{ICSE}}. \bibinfo{pages}{2374--2386}.
\newblock


\bibitem[Johnson(2016)]%
        {johnson2016itch}
\bibfield{author}{\bibinfo{person}{David~E. Johnson}.} \bibinfo{year}{2016}\natexlab{}.
\newblock \showarticletitle{ITCH: Individual Testing of Computer Homework for Scratch Assignments}. In \bibinfo{booktitle}{\emph{SIGCSE}}. \bibinfo{pages}{223--227}.
\newblock
\href{https://doi.org/10.1145/2839509.2844600}{doi:\nolinkurl{10.1145/2839509.2844600}}


\bibitem[Keuning et~al\mbox{.}(2019)]%
        {keuning2019automatedfeedback}
\bibfield{author}{\bibinfo{person}{Hieke Keuning}, \bibinfo{person}{Johan Jeuring}, {and} \bibinfo{person}{Bastiaan Heeren}.} \bibinfo{year}{2019}\natexlab{}.
\newblock \showarticletitle{A Systematic Literature Review of Automated Feedback Generation for Programming Exercises}.
\newblock \bibinfo{journal}{\emph{ACM Transactions on Computing Education}} \bibinfo{volume}{19}, \bibinfo{number}{1} (\bibinfo{year}{2019}), \bibinfo{pages}{3:1--3:43}.
\newblock
\href{https://doi.org/10.1145/3231711}{doi:\nolinkurl{10.1145/3231711}}


\bibitem[Ko and Myers(2004)]%
        {ko2004whyline}
\bibfield{author}{\bibinfo{person}{Andrew~J. Ko} {and} \bibinfo{person}{Brad~A. Myers}.} \bibinfo{year}{2004}\natexlab{}.
\newblock \showarticletitle{Designing the Whyline: A Debugging Interface for Asking Questions about Program Behavior}. In \bibinfo{booktitle}{\emph{CHI}}. \bibinfo{pages}{151--158}.
\newblock
\href{https://doi.org/10.1145/985692.985712}{doi:\nolinkurl{10.1145/985692.985712}}


\bibitem[Liffiton et~al\mbox{.}(2023)]%
        {liffiton2023codehelp}
\bibfield{author}{\bibinfo{person}{Mark Liffiton}, \bibinfo{person}{Brad Sheese}, \bibinfo{person}{Jaromir Savelka}, {and} \bibinfo{person}{Paul Denny}.} \bibinfo{year}{2023}\natexlab{}.
\newblock \showarticletitle{CodeHelp: Using Large Language Models with Guardrails for Scalable Support in Programming Classes}.
\newblock \bibinfo{journal}{\emph{Proceedings of the 23rd International Conference on Computing Education Research (Koli Calling 2023)}} (\bibinfo{year}{2023}).
\newblock
\href{https://doi.org/10.1145/3631802.3631830}{doi:\nolinkurl{10.1145/3631802.3631830}}


\bibitem[Maloney et~al\mbox{.}(2010)]%
        {maloney2010scratch}
\bibfield{author}{\bibinfo{person}{John Maloney}, \bibinfo{person}{Mitchel Resnick}, \bibinfo{person}{Natalie Rusk}, \bibinfo{person}{Brian Silverman}, {and} \bibinfo{person}{Evelyn Eastmond}.} \bibinfo{year}{2010}\natexlab{}.
\newblock \showarticletitle{The scratch programming language and environment}.
\newblock \bibinfo{journal}{\emph{ACM Transactions on Computing Education (TOCE)}} \bibinfo{volume}{10}, \bibinfo{number}{4} (\bibinfo{year}{2010}), \bibinfo{pages}{1--15}.
\newblock


\bibitem[Marwan and Price(2023)]%
        {marwan_2023_isnap_tlt}
\bibfield{author}{\bibinfo{person}{Samiha Marwan} {and} \bibinfo{person}{Thomas~W. Price}.} \bibinfo{year}{2023}\natexlab{}.
\newblock \showarticletitle{iSnap: Evolution and Evaluation of a Data-Driven Hint System for Block-Based Programming}.
\newblock \bibinfo{journal}{\emph{IEEE Transactions on Learning Technologies}} \bibinfo{volume}{16}, \bibinfo{number}{3} (\bibinfo{year}{2023}), \bibinfo{pages}{399--413}.
\newblock
\href{https://doi.org/10.1109/TLT.2022.3223577}{doi:\nolinkurl{10.1109/TLT.2022.3223577}}


\bibitem[Mens and Tourw{\'e}(2004)]%
        {Mens2004SurveyRefactoring}
\bibfield{author}{\bibinfo{person}{Tom Mens} {and} \bibinfo{person}{Tom Tourw{\'e}}.} \bibinfo{year}{2004}\natexlab{}.
\newblock \showarticletitle{A Survey of Software Refactoring}.
\newblock \bibinfo{journal}{\emph{IEEE Transactions on Software Engineering}} \bibinfo{volume}{30}, \bibinfo{number}{2} (\bibinfo{year}{2004}), \bibinfo{pages}{126--139}.
\newblock
\href{https://doi.org/10.1109/TSE.2004.1265817}{doi:\nolinkurl{10.1109/TSE.2004.1265817}}


\bibitem[Moreno-Le{\'o}n and Robles(2015a)]%
        {moreno2015drscratch}
\bibfield{author}{\bibinfo{person}{Jes{\'u}s Moreno-Le{\'o}n} {and} \bibinfo{person}{Gregorio Robles}.} \bibinfo{year}{2015}\natexlab{a}.
\newblock \showarticletitle{Dr. Scratch: A Web Tool to Automatically Evaluate Scratch Projects}. In \bibinfo{booktitle}{\emph{WiPSCE}}. \bibinfo{pages}{132--133}.
\newblock
\href{https://doi.org/10.1145/2818314.2818338}{doi:\nolinkurl{10.1145/2818314.2818338}}


\bibitem[Moreno-Le{\'o}n and Robles(2015b)]%
        {moreno_2015_drscratch}
\bibfield{author}{\bibinfo{person}{Jes{\'u}s Moreno-Le{\'o}n} {and} \bibinfo{person}{Gregorio Robles}.} \bibinfo{year}{2015}\natexlab{b}.
\newblock \showarticletitle{Dr. Scratch: Automatic Analysis of Scratch Projects to Assess and Foster Computational Thinking}. In \bibinfo{booktitle}{\emph{Proceedings of the Workshop in Primary and Secondary Computing Education (WiPSCE)}}.
\newblock
\href{https://doi.org/10.1145/2818314.2818338}{doi:\nolinkurl{10.1145/2818314.2818338}}


\bibitem[Oberm{\"u}ller and Fraser(2024)]%
        {obermueller2024fixes}
\bibfield{author}{\bibinfo{person}{Florian Oberm{\"u}ller} {and} \bibinfo{person}{Gordon Fraser}.} \bibinfo{year}{2024}\natexlab{}.
\newblock \showarticletitle{Do Scratchers Fix Their Bugs? Detecting Fixes of Scratch Static Analysis Warnings}. In \bibinfo{booktitle}{\emph{Proceedings of the 19th Workshop in Primary and Secondary Computing Education (WiPSCE '24)}}. \bibinfo{publisher}{ACM}, \bibinfo{pages}{18:1--18:4}.
\newblock
\href{https://doi.org/10.1145/3677619.3678108}{doi:\nolinkurl{10.1145/3677619.3678108}}


\bibitem[Oberm{\"u}ller et~al\mbox{.}(2023)]%
        {obermuller2023effects}
\bibfield{author}{\bibinfo{person}{Florian Oberm{\"u}ller}, \bibinfo{person}{Luisa Greifenstein}, {and} \bibinfo{person}{Gordon Fraser}.} \bibinfo{year}{2023}\natexlab{}.
\newblock \showarticletitle{Effects of Automated Feedback in Scratch Programming Tutorials}. In \bibinfo{booktitle}{\emph{Proceedings of the 2023 ACM Conference on Innovation and Technology in Computer Science Education (ITiCSE '23)}}. \bibinfo{publisher}{ACM}, \bibinfo{pages}{396--402}.
\newblock
\href{https://doi.org/10.1145/3587102.3588803}{doi:\nolinkurl{10.1145/3587102.3588803}}


\bibitem[Oberm{\"u}ller et~al\mbox{.}(2021)]%
        {obermuller2021catnip}
\bibfield{author}{\bibinfo{person}{Florian Oberm{\"u}ller}, \bibinfo{person}{Ute Heuer}, {and} \bibinfo{person}{Gordon Fraser}.} \bibinfo{year}{2021}\natexlab{}.
\newblock \showarticletitle{Guiding Next-Step Hint Generation Using Automated Tests}. In \bibinfo{booktitle}{\emph{Proceedings of the 26th ACM Conference on Innovation and Technology in Computer Science Education (ITiCSE '21)}}. \bibinfo{publisher}{ACM}, \bibinfo{pages}{220--226}.
\newblock
\href{https://doi.org/10.1145/3430665.3456344}{doi:\nolinkurl{10.1145/3430665.3456344}}


\bibitem[Opdyke(1992)]%
        {Opdyke1992RefactoringFrameworks}
\bibfield{author}{\bibinfo{person}{William~F. Opdyke}.} \bibinfo{year}{1992}\natexlab{}.
\newblock \emph{\bibinfo{title}{Refactoring Object-Oriented Frameworks}}.
\newblock \bibinfo{thesistype}{Ph.\,D. Dissertation}. \bibinfo{school}{University of Illinois at Urbana–Champaign}, \bibinfo{address}{Urbana, IL, USA}.
\newblock
\urldef\tempurl%
\url{http://hdl.handle.net/2142/72072}
\showURL{%
\tempurl}


\bibitem[{Open Logic Project}(nd)]%
        {OpenLogicAlphaConversion}
\bibfield{author}{\bibinfo{person}{{Open Logic Project}}.} \bibinfo{year}{n.d.}\natexlab{}.
\newblock \bibinfo{title}{${\\alpha}$-Conversion}.
\newblock \bibinfo{howpublished}{The Open Logic Text}.
\newblock
\urldef\tempurl%
\url{https://builds.openlogicproject.org/content/lambda-calculus/syntax/alpha.pdf}
\showURL{%
\tempurl}
\newblock
\shownote{Accessed 2025-09-29}.


\bibitem[Price et~al\mbox{.}(2017a)]%
        {price_aied_2017_hint_quality}
\bibfield{author}{\bibinfo{person}{Thomas~W. Price}, \bibinfo{person}{Yihuan Dong}, {and} \bibinfo{person}{Dragan Lipovac}.} \bibinfo{year}{2017}\natexlab{a}.
\newblock \showarticletitle{Hint Generation Under Uncertainty: The Effect of Hint Quality on Help-Seeking Behavior}. In \bibinfo{booktitle}{\emph{Proceedings of AIED 2017}}.
\newblock
\urldef\tempurl%
\url{https://isnap.csc.ncsu.edu/home/public/papers/PriceAIED2017.pdf}
\showURL{%
\tempurl}


\bibitem[Price et~al\mbox{.}(2017b)]%
        {price2017isnap}
\bibfield{author}{\bibinfo{person}{Thomas~W. Price}, \bibinfo{person}{Yihuan Dong}, {and} \bibinfo{person}{Dragan Lipovac}.} \bibinfo{year}{2017}\natexlab{b}.
\newblock \showarticletitle{iSnap: Towards Intelligent Tutoring in Novice Programming Environments}. In \bibinfo{booktitle}{\emph{SIGCSE}}. \bibinfo{pages}{483--488}.
\newblock
\href{https://doi.org/10.1145/3017680.3017762}{doi:\nolinkurl{10.1145/3017680.3017762}}


\bibitem[Price et~al\mbox{.}(2017c)]%
        {price_2017_isnap_sigcse}
\bibfield{author}{\bibinfo{person}{Thomas~W. Price}, \bibinfo{person}{Yihuan Dong}, {and} \bibinfo{person}{Dragan Lipovac}.} \bibinfo{year}{2017}\natexlab{c}.
\newblock \showarticletitle{iSnap: Towards Intelligent Tutoring in Novice Programming Environments}. In \bibinfo{booktitle}{\emph{Proceedings of SIGCSE 2017}}. \bibinfo{pages}{483--488}.
\newblock
\href{https://doi.org/10.1145/3017680.3017762}{doi:\nolinkurl{10.1145/3017680.3017762}}


\bibitem[Price et~al\mbox{.}(2019)]%
        {price2019comparison}
\bibfield{author}{\bibinfo{person}{Thomas~W Price}, \bibinfo{person}{Yihuan Dong}, \bibinfo{person}{Rui Zhi}, \bibinfo{person}{Benjamin Paa{\ss}en}, \bibinfo{person}{Nicholas Lytle}, \bibinfo{person}{Veronica Catet{\'e}}, {and} \bibinfo{person}{Tiffany Barnes}.} \bibinfo{year}{2019}\natexlab{}.
\newblock \showarticletitle{A comparison of the quality of data-driven programming hint generation algorithms}.
\newblock \bibinfo{journal}{\emph{International Journal of Artificial Intelligence in Education}} \bibinfo{volume}{29}, \bibinfo{number}{3} (\bibinfo{year}{2019}), \bibinfo{pages}{368--395}.
\newblock


\bibitem[Resnick et~al\mbox{.}(2009)]%
        {resnick2009scratch}
\bibfield{author}{\bibinfo{person}{Mitchel Resnick}, \bibinfo{person}{John Maloney}, \bibinfo{person}{Andr{\'e}s Monroy-Hern{\'a}ndez}, \bibinfo{person}{Natalie Rusk}, \bibinfo{person}{Evelyn Eastmond}, \bibinfo{person}{Karen Brennan}, \bibinfo{person}{Amon Millner}, \bibinfo{person}{Eric Rosenbaum}, \bibinfo{person}{Jay Silver}, \bibinfo{person}{Brian Silverman}, {et~al\mbox{.}}} \bibinfo{year}{2009}\natexlab{}.
\newblock \showarticletitle{Scratch: programming for all}.
\newblock \bibinfo{journal}{\emph{Commun. ACM}} \bibinfo{volume}{52}, \bibinfo{number}{11} (\bibinfo{year}{2009}), \bibinfo{pages}{60--67}.
\newblock


\bibitem[Richland et~al\mbox{.}(2009)]%
        {richland2009pretesting}
\bibfield{author}{\bibinfo{person}{Lindsey~E. Richland}, \bibinfo{person}{Nate Kornell}, {and} \bibinfo{person}{Liche~Sean Kao}.} \bibinfo{year}{2009}\natexlab{}.
\newblock \showarticletitle{The Pretesting Effect: Do Unsuccessful Retrieval Attempts Enhance Learning?}
\newblock \bibinfo{journal}{\emph{Journal of Experimental Psychology: Applied}} \bibinfo{volume}{15}, \bibinfo{number}{3} (\bibinfo{year}{2009}), \bibinfo{pages}{243--257}.
\newblock
\href{https://doi.org/10.1037/a0016496}{doi:\nolinkurl{10.1037/a0016496}}


\bibitem[Rivers and Koedinger(2017)]%
        {rivers2017itap}
\bibfield{author}{\bibinfo{person}{Kelly Rivers} {and} \bibinfo{person}{Kenneth~R. Koedinger}.} \bibinfo{year}{2017}\natexlab{}.
\newblock \showarticletitle{Data-Driven Hint Generation in Vast Solution Spaces: A Self-Improving Python Programming Tutor}.
\newblock \bibinfo{journal}{\emph{International Journal of Artificial Intelligence in Education}} \bibinfo{volume}{27}, \bibinfo{number}{1} (\bibinfo{year}{2017}), \bibinfo{pages}{37--64}.
\newblock
\href{https://doi.org/10.1007/s40593-015-0070-z}{doi:\nolinkurl{10.1007/s40593-015-0070-z}}


\bibitem[Roy and Cordy(2007)]%
        {RoyCordy2007CloneSurvey}
\bibfield{author}{\bibinfo{person}{Chanchal~K. Roy} {and} \bibinfo{person}{James~R. Cordy}.} \bibinfo{year}{2007}\natexlab{}.
\newblock \bibinfo{booktitle}{\emph{A Survey on Software Clone Detection Research}}.
\newblock \bibinfo{type}{{T}echnical {R}eport} Technical Report 2007-541. \bibinfo{institution}{School of Computing, Queen’s University}, \bibinfo{address}{Kingston, Ontario, Canada}.
\newblock
\urldef\tempurl%
\url{https://research.cs.queensu.ca/TechReports/Reports/2007-541.pdf}
\showURL{%
\tempurl}


\bibitem[Santolucito et~al\mbox{.}(2022)]%
        {Santolucito22}
\bibfield{author}{\bibinfo{person}{Mark Santolucito}, \bibinfo{person}{Jialu Zhang}, \bibinfo{person}{Ennan Zhai}, \bibinfo{person}{Jürgen Cito}, {and} \bibinfo{person}{Ruzica Piskac}.} \bibinfo{year}{2022}\natexlab{}.
\newblock \showarticletitle{Learning CI Configuration Correctness for Early Build Feedback}. In \bibinfo{booktitle}{\emph{2022 IEEE International Conference on Software Analysis, Evolution and Reengineering (SANER)}}. \bibinfo{pages}{1006--1017}.
\newblock
\href{https://doi.org/10.1109/SANER53432.2022.00118}{doi:\nolinkurl{10.1109/SANER53432.2022.00118}}


\bibitem[Schweikl and Fraser(2025)]%
        {schweikl2025repurr}
\bibfield{author}{\bibinfo{person}{Sebastian Schweikl} {and} \bibinfo{person}{Gordon Fraser}.} \bibinfo{year}{2025}\natexlab{}.
\newblock \showarticletitle{RePurr: Automated Repair of Block-Based Learners’ Programs}.
\newblock \bibinfo{journal}{\emph{Proceedings of the ACM on Software Engineering}} \bibinfo{volume}{2}, \bibinfo{number}{FSE} (\bibinfo{year}{2025}), \bibinfo{pages}{1475--1498}.
\newblock


\bibitem[Shao et~al\mbox{.}(2025)]%
        {shao2025llmscorrectlyintegratedsoftware}
\bibfield{author}{\bibinfo{person}{Yuchen Shao}, \bibinfo{person}{Yuheng Huang}, \bibinfo{person}{Jiawei Shen}, \bibinfo{person}{Lei Ma}, \bibinfo{person}{Ting Su}, {and} \bibinfo{person}{Chengcheng Wan}.} \bibinfo{year}{2025}\natexlab{}.
\newblock \bibinfo{title}{Are LLMs Correctly Integrated into Software Systems?}
\newblock
\showeprint[arxiv]{2407.05138}~[cs.SE]
\urldef\tempurl%
\url{https://arxiv.org/abs/2407.05138}
\showURL{%
\tempurl}


\bibitem[Si et~al\mbox{.}(2025)]%
        {si2025viscratchusinglargelanguage}
\bibfield{author}{\bibinfo{person}{Yuan Si}, \bibinfo{person}{Daming Li}, \bibinfo{person}{Hanyuan Shi}, {and} \bibinfo{person}{Jialu Zhang}.} \bibinfo{year}{2025}\natexlab{}.
\newblock \bibinfo{title}{ViScratch: Using Large Language Models and Gameplay Videos for Automated Feedback in Scratch}.
\newblock
\showeprint[arxiv]{2509.11065}~[cs.SE]
\urldef\tempurl%
\url{https://arxiv.org/abs/2509.11065}
\showURL{%
\tempurl}


\bibitem[Stahlbauer et~al\mbox{.}(2019)]%
        {stahlbauer2019whisker}
\bibfield{author}{\bibinfo{person}{Andreas Stahlbauer}, \bibinfo{person}{Michael Kreis}, {and} \bibinfo{person}{Gordon Fraser}.} \bibinfo{year}{2019}\natexlab{}.
\newblock \showarticletitle{Testing Scratch Programs Automatically}. In \bibinfo{booktitle}{\emph{ESEC/FSE}}. \bibinfo{pages}{165--175}.
\newblock


\bibitem[Strijbol et~al\mbox{.}(2024)]%
        {STRIJBOL2024101617}
\bibfield{author}{\bibinfo{person}{Niko Strijbol}, \bibinfo{person}{Robbe {De Proft}}, \bibinfo{person}{Klaas Goethals}, \bibinfo{person}{Bart Mesuere}, \bibinfo{person}{Peter Dawyndt}, {and} \bibinfo{person}{Christophe Scholliers}.} \bibinfo{year}{2024}\natexlab{}.
\newblock \showarticletitle{Blink: An educational software debugger for Scratch}.
\newblock \bibinfo{journal}{\emph{SoftwareX}}  \bibinfo{volume}{25} (\bibinfo{year}{2024}), \bibinfo{pages}{101617}.
\newblock
\showISSN{2352-7110}
\href{https://doi.org/10.1016/j.softx.2023.101617}{doi:\nolinkurl{10.1016/j.softx.2023.101617}}


\bibitem[Techapalokul and Tilevich(2017)]%
        {techapalokul2017qualityhound}
\bibfield{author}{\bibinfo{person}{Peeratham Techapalokul} {and} \bibinfo{person}{Eli Tilevich}.} \bibinfo{year}{2017}\natexlab{}.
\newblock \showarticletitle{Quality Hound—An Online Code Smell Analyzer for Scratch Programs}. In \bibinfo{booktitle}{\emph{IEEE VL/HCC}}. \bibinfo{pages}{277--281}.
\newblock


\bibitem[van~de Pol et~al\mbox{.}(2010)]%
        {vandePol2010Scaffolding}
\bibfield{author}{\bibinfo{person}{Janneke van~de Pol}, \bibinfo{person}{Monique Volman}, {and} \bibinfo{person}{Jos Beishuizen}.} \bibinfo{year}{2010}\natexlab{}.
\newblock \showarticletitle{Scaffolding in Teacher--Student Interaction: A Decade of Research}.
\newblock \bibinfo{journal}{\emph{Educational Psychology Review}} \bibinfo{volume}{22}, \bibinfo{number}{3} (\bibinfo{year}{2010}), \bibinfo{pages}{271--296}.
\newblock
\href{https://doi.org/10.1007/s10648-010-9127-6}{doi:\nolinkurl{10.1007/s10648-010-9127-6}}


\bibitem[Wang et~al\mbox{.}(2021)]%
        {wang2021snapcheck}
\bibfield{author}{\bibinfo{person}{Wengran Wang}, \bibinfo{person}{Chenhao Zhang}, \bibinfo{person}{Andreas Stahlbauer}, \bibinfo{person}{Gordon Fraser}, {and} \bibinfo{person}{Thomas~W. Price}.} \bibinfo{year}{2021}\natexlab{}.
\newblock \showarticletitle{SnapCheck: Automated Testing for Snap! Programs}. In \bibinfo{booktitle}{\emph{Proceedings of the 26th ACM Conference on Innovation and Technology in Computer Science Education (ITiCSE '21)}}. \bibinfo{publisher}{ACM}, \bibinfo{pages}{227--233}.
\newblock
\href{https://doi.org/10.1145/3430665.3456367}{doi:\nolinkurl{10.1145/3430665.3456367}}


\bibitem[Wiggins et~al\mbox{.}(2021)]%
        {wiggins_2021_hint_requests}
\bibfield{author}{\bibinfo{person}{Joseph~B. Wiggins}, \bibinfo{person}{Fahmid~M. Fahid}, \bibinfo{person}{Andrew Emerson}, \bibinfo{person}{Madeline Hinckle}, \bibinfo{person}{Andy Smith}, \bibinfo{person}{Kristy~E. Boyer}, \bibinfo{person}{Bradford~W. Mott}, \bibinfo{person}{Eric~N. Wiebe}, {and} \bibinfo{person}{James~C. Lester}.} \bibinfo{year}{2021}\natexlab{}.
\newblock \showarticletitle{Exploring Novice Programmers' Hint Requests in an Intelligent Block-Based Coding Environment}. In \bibinfo{booktitle}{\emph{Proceedings of SIGCSE 2021}}. \bibinfo{pages}{52--58}.
\newblock
\href{https://doi.org/10.1145/3408877.3432538}{doi:\nolinkurl{10.1145/3408877.3432538}}


\bibitem[Wood et~al\mbox{.}(1976)]%
        {WoodBrunerRoss1976}
\bibfield{author}{\bibinfo{person}{David~J. Wood}, \bibinfo{person}{Jerome~S. Bruner}, {and} \bibinfo{person}{Gail Ross}.} \bibinfo{year}{1976}\natexlab{}.
\newblock \showarticletitle{The Role of Tutoring in Problem Solving}.
\newblock \bibinfo{journal}{\emph{Journal of Child Psychology and Psychiatry}} \bibinfo{volume}{17}, \bibinfo{number}{2} (\bibinfo{year}{1976}), \bibinfo{pages}{89--100}.
\newblock
\href{https://doi.org/10.1111/j.1469-7610.1976.tb00381.x}{doi:\nolinkurl{10.1111/j.1469-7610.1976.tb00381.x}}


\bibitem[Zhang et~al\mbox{.}(2024)]%
        {PyDex}
\bibfield{author}{\bibinfo{person}{Jialu Zhang}, \bibinfo{person}{Jos{\'{e}}~Pablo Cambronero}, \bibinfo{person}{Sumit Gulwani}, \bibinfo{person}{Vu Le}, \bibinfo{person}{Ruzica Piskac}, \bibinfo{person}{Gustavo Soares}, {and} \bibinfo{person}{Gust Verbruggen}.} \bibinfo{year}{2024}\natexlab{}.
\newblock \showarticletitle{PyDex: Repairing Bugs in Introductory Python Assignments using LLMs}.
\newblock \bibinfo{journal}{\emph{Proc. {ACM} Program. Lang.}} \bibinfo{volume}{8}, \bibinfo{number}{{OOPSLA1}} (\bibinfo{year}{2024}), \bibinfo{pages}{1100--1124}.
\newblock
\href{https://doi.org/10.1145/3649850}{doi:\nolinkurl{10.1145/3649850}}


\bibitem[Zhang et~al\mbox{.}(2025)]%
        {zhang2025systematicstudytimelimit}
\bibfield{author}{\bibinfo{person}{Jialu Zhang}, \bibinfo{person}{Jialiang Gu}, \bibinfo{person}{Wangmeiyu Zhang}, \bibinfo{person}{José~Pablo Cambronero}, \bibinfo{person}{John Kolesar}, \bibinfo{person}{Ruzica Piskac}, \bibinfo{person}{Daming Li}, {and} \bibinfo{person}{Hanyuan Shi}.} \bibinfo{year}{2025}\natexlab{}.
\newblock \bibinfo{title}{A Systematic Study of Time Limit Exceeded Errors in Online Programming Assignments}.
\newblock
\showeprint[arxiv]{2510.14339}~[cs.SE]
\urldef\tempurl%
\url{https://arxiv.org/abs/2510.14339}
\showURL{%
\tempurl}


\bibitem[Zhang et~al\mbox{.}(2023)]%
        {clef}
\bibfield{author}{\bibinfo{person}{Jialu Zhang}, \bibinfo{person}{De Li}, \bibinfo{person}{John~Charles Kolesar}, \bibinfo{person}{Hanyuan Shi}, {and} \bibinfo{person}{Ruzica Piskac}.} \bibinfo{year}{2023}\natexlab{}.
\newblock \showarticletitle{Automated Feedback Generation for Competition-Level Code}. In \bibinfo{booktitle}{\emph{Proceedings of the 37th IEEE/ACM International Conference on Automated Software Engineering}} (Rochester, MI, USA) \emph{(\bibinfo{series}{ASE '22})}. \bibinfo{publisher}{Association for Computing Machinery}, \bibinfo{address}{New York, NY, USA}, Article \bibinfo{articleno}{13}, \bibinfo{numpages}{13}~pages.
\newblock
\showISBNx{9781450394758}
\href{https://doi.org/10.1145/3551349.3560425}{doi:\nolinkurl{10.1145/3551349.3560425}}


\bibitem[Zhang et~al\mbox{.}(2022)]%
        {gmerge}
\bibfield{author}{\bibinfo{person}{Jialu Zhang}, \bibinfo{person}{Todd Mytkowicz}, \bibinfo{person}{Mike Kaufman}, \bibinfo{person}{Ruzica Piskac}, {and} \bibinfo{person}{Shuvendu~K. Lahiri}.} \bibinfo{year}{2022}\natexlab{}.
\newblock \showarticletitle{Using Pre-Trained Language Models to Resolve Textual and Semantic Merge Conflicts (Experience Paper)}. In \bibinfo{booktitle}{\emph{Proceedings of the 31st ACM SIGSOFT International Symposium on Software Testing and Analysis}} (Virtual, South Korea) \emph{(\bibinfo{series}{ISSTA 2022})}. \bibinfo{publisher}{Association for Computing Machinery}, \bibinfo{address}{New York, NY, USA}, \bibinfo{pages}{77–88}.
\newblock
\showISBNx{9781450393799}
\href{https://doi.org/10.1145/3533767.3534396}{doi:\nolinkurl{10.1145/3533767.3534396}}


\bibitem[Zhang et~al\mbox{.}(2021)]%
        {ConfigX}
\bibfield{author}{\bibinfo{person}{Jialu Zhang}, \bibinfo{person}{Ruzica Piskac}, \bibinfo{person}{Ennan Zhai}, {and} \bibinfo{person}{Tianyin Xu}.} \bibinfo{year}{2021}\natexlab{}.
\newblock \showarticletitle{Static detection of silent misconfigurations with deep interaction analysis}.
\newblock \bibinfo{journal}{\emph{Proc. ACM Program. Lang.}} \bibinfo{volume}{5}, \bibinfo{number}{OOPSLA}, Article \bibinfo{articleno}{140} (\bibinfo{date}{Oct.} \bibinfo{year}{2021}), \bibinfo{numpages}{30}~pages.
\newblock
\href{https://doi.org/10.1145/3485517}{doi:\nolinkurl{10.1145/3485517}}


\end{thebibliography}

\end{document}